\documentclass[aip,jcp, reprint,groupedaddress]{revtex4-2}
\usepackage{graphicx}
\usepackage{tabularx}
\usepackage{dcolumn}
\usepackage{longtable}
\usepackage{tensor}
\usepackage{color}
\usepackage{placeins}
\usepackage{bm}
\usepackage{amsmath}
\usepackage{amsfonts}
\usepackage{amssymb}
\usepackage{float}
\usepackage[urlcolor=white]{hyperref}%
\providecommand{\U}[1]{\protect\rule{.1in}{.1in}}
\graphicspath{{./Figures/}{"C:/Users/GROUP
LCPT/Documents/Davide/SHA_geometric_properties/Figures/"
}}

\begin{document}
\title{On the single-Hessian Gaussian wavepacket dynamics}

\author{Davide Barbiero}
\author{Ji\v{r}\'i J. L. Van\'i\v{c}ek}
\email{jiri.vanicek@epfl.ch}

\affiliation{Laboratory of Theoretical Physical Chemistry, Institut des Sciences et Ing\'enierie Chimiques, Ecole Polytechnique F\'ed\'erale de Lausanne (EPFL), CH-1015, Lausanne, Switzerland}

\date{\today}

\begin{abstract}
Single-Hessian Gaussian wavepacket dynamics (GWD) significantly reduces the computational burden of Heller's local harmonic GWD, while maintaining comparable accuracy in approximating vibronic spectra. 
Here, we provide a new, symplectic derivation of the equations of motion of single-Hessian GWD and show that, unlike the local harmonic version, this method conserves the non-canonical symplectic structure on the manifold of Gaussian wavepackets and--for bounded dynamics in smooth potentials--avoids the drift of energy.
Our numerical results suggest that, despite being much more efficient than the local harmonic variant, the single-Hessian GWD exhibits the same $\mathcal{O}(\hbar)$ asymptotic error in averages of observables. To further accelerate numerical simulations, we implement high-order time-stepping geometric integrators that are time-reversible and conserve the norm and symplectic structure exactly, regardless of the time step. 
In addition, we present explicit expressions for the exact evolution of the width of a single-Hessian Gaussian wavepacket in a general potential, as well as for the exact evolution of the whole wavepacket in a global harmonic potential.
Using on-the-fly ab initio Gaussian wavepacket dynamics on the first excited-state surface of ammonia, we numerically confirm the conservation of geometric properties by these integrators and demonstrate that high-order integrators can enhance both accuracy and computational efficiency. We also compute the photoelectron spectrum of the difluorocarbene anion and the absorption spectrum of methylamine, and find that, in comparison with experiment, single-Hessian GWD outperforms global harmonic models and matches the accuracy of local harmonic GWD. Finally, we identify which spectral features are sensitive to the choice of reference Hessian.
\end{abstract}

\maketitle


\section{\label{sec:introduction}Introduction}

Vibrationally resolved electronic spectra play a pivotal role in physical chemistry, providing detailed insight into molecular structure and dynamics.~\cite{Herzberg:1966,Quack_Merkt:2011_v2,book_Heller:2018} Simulating these spectra requires a quantum-mechanical treatment, yet the exponential scaling of numerically exact approaches restricts their application to small systems.~\cite{book_Lubich:2008,Choi_Vanicek:2021}

To overcome the curse of dimensionality in large molecules, approximations are essential.
A popular approach involves global harmonic models,~\cite{AvilaFerrer_Santoro:2012,Baiardi_Barone:2013} for which an exact, analytical solution to the Schr\"{o}dinger equation is known.
However, anharmonicity can affect vibronic spectra significantly. Although many quantum~\cite{Worth_Burghardt:2003,Luis_Christiansen:2006,Petrenko_Rauhut:2017} and semiclassical~\cite{Walton_Manolopoulos:1995,Ben-Nun_Martinez:1999,Tatchen_Pollak:2009,Sulc_Vanicek:2013} methods include anharmonic effects, they often entail high computational costs.
A notable exception is the local harmonic Gaussian wavepacket dynamics (GWD),~\cite{Heller:1975} a single-trajectory semiclassical approach that has proven sufficiently accurate for calculating low- and medium-resolution vibronic spectra of molecules with weakly anharmonic potential energy surfaces (PESs).~\cite{Wehrle_Vanicek:2014,Wehrle_Vanicek:2015,Kletnieks_Vanicek:2023}
The success of this crude approximation reflects the forgiving nature of spectra, whose main features are often insensitive to fine details of the wavefunction. Instead, low- to medium-resolution spectra typically depend on broader dynamical properties that can be reproduced even by approximate methods. For example, although local harmonic GWD does not capture tunnelling, wavepacket splitting, or multi-path interference, this method can still provide reliable spectra when such processes occur on timescales longer than those associated with external broadening.~\cite{Begusic_Vanicek:2022}
More recently, the local harmonic method has been generalized to capture nonzero-temperature~\cite{Begusic_Vanicek:2020} and Herzberg-Teller~\cite{Patoz_Vanicek:2018,Begusic_Vanicek:2018} effects, and applied to simulate ultrafast nonlinear spectra~\cite{Begusic_Vanicek:2018a_v2,Begusic_Vanicek:2020a} and internal conversion rates.~\cite{Wenzel_Mitric:2023,Wenzel_Mitric:2023a_v2} 

Despite its success, the local harmonic GWD can become computationally expensive when combined with on-the-fly ab initio evaluation of the electronic structure. Constructing a local harmonic potential about the center of the evolving wavepacket requires evaluating the Hessian matrix at each time step, an approach that is often infeasible for large systems or when high-level electronic structure methods are used, particularly for excited-state dynamics.
Because this issue is ubiquitous in (semiclassical) ab initio molecular dynamics,~\cite{Tatchen_Miller:2011,Rognoni_Ceotto:2021} numerous strategies have been devised to mitigate it.
A non-exhaustive list includes (i)~Hessian updating schemes, where the Hessian is approximated in a step-wise fashion,~\cite{Bakken_Schlegel:1999,Ceotto_Hase:2013} (ii)~Hessian interpolation schemes, where the Hessian is computed only at selected points along the classical trajectory and interpolated in between,~\cite{Wehrle_Vanicek:2014,Begusic_Vanicek:2020} (iii)~clustering of similar configurations via machine learning algorithms and assigning a unique Hessian to each cluster,~\cite{Gandolfi_Ceotto:2021} and (iv)~analytical differentiation of local Gaussian-regression fits to the PES.~\cite{Laude_Richardson:2018,Gherib_Genin:2024_v2}

To reduce the computational cost of the local harmonic GWD more substantially, Begu\v{s}i\'{c}, Cordova, and Van\'i\v{c}ek proposed the single-Hessian GWD,~\cite{Begusic_Vanicek:2019} where the center of the wavepacket still follows the exact anharmonic classical trajectory, but the width is propagated using a constant Hessian.
The single-Hessian GWD has allowed the computation of the vibronic spectra of several molecules at the cost of a single classical trajectory, while achieving accuracy comparable to the accuracy of the much more expensive local harmonic GWD.~\cite{Begusic_Vanicek:2019,Prlj_Vanicek:2020,Begusic_Vanicek:2021}
The efficiency and unexpected accuracy of the single-Hessian approach--attributed to its superior geometric properties over the local harmonic version~\cite{Vanicek:2023}--have motivated its implementation in the electronic structure package Turbomole.~\cite{Begusic_Vanicek:2022,Franzke_Weigend:2023}

In a sequence of recent mathematical breakthroughs,~\cite{Lasser_Lubich:2020,Ohsawa:2021,Burkhard_Lasser:2024} it was shown that observables may converge faster than the wavefunction in GWD. Whereas the wavefunction in local harmonic, symplectic, and variational GWD has the same $L^{2}$-norm error, namely error of order $\mathcal{O}(\hbar^{1/2})$ in terms of the semiclassical parameter $\hbar$,~\cite{Hagedorn:1980,book_Lubich:2008,Ohsawa:2021} the expectation values of observables have been shown to have at least $\mathcal{O}(\hbar)$ error in local harmonic GWD,~\cite{Ohsawa:2021} $\mathcal{O}(\hbar^{3/2})$ error [$\mathcal{O}(\hbar^2)$  for energy] in symplectic GWD,~\cite{Ohsawa:2021} and $\mathcal{O}(\hbar^{2})$ error in variational GWD.~\cite{Burkhard_Lasser:2024} 
This raises the concern that the more efficient single-Hessian GWD may suffer from a reduced order of accuracy, potentially as low as $\mathcal{O}(\hbar^{0})$.

Our goal is to analyze the single-Hessian GWD in detail, especially its accuracy and geometric properties, introduce high-order symplectic integrators designed to accelerate on-the-fly numerical simulations, and apply the method to new molecular systems. 
In Sec.~\ref{sec:theory}, we (i)~review the single-Hessian approximation of the exact potential, (ii)~present an alternative, ``symplectic'' derivation of the equations of motion (which directly implies the conservation of symplectic structure and effective energy), and (iii)~examine the geometric properties of the single-Hessian GWD.
In Sec.~\ref{sec:integrators}, we describe efficient geometric time-stepping integrators that preserve most of these properties exactly. We also provide exact integrators for the evolution of the width of a single-Hessian Gaussian wavepacket in any potential and for the evolution of all parameters of a Gaussian wavepacket in a global harmonic potential.
Sections~\ref{sec:details} and~\ref{sec:results} provide numerical examples that demonstrate the improved properties of the single-Hessian GWD compared to the original local harmonic method and confirm the fast convergence, increased efficiency, and preservation of the geometric properties by the high-order integrators in on-the-fly ab initio simulations. 
In addition, we numerically explore the accuracy of single-Hessian, local harmonic, and variational GWD with respect to the semiclassical parameter $\hbar$.
To highlight the benefits and limitations of the method, we compare spectra of four different diatomic molecules computed using single-Hessian GWD and second-order perturbation theory, a widely used approach for incorporating anharmonic effects in spectroscopy calculations.
Finally, we use on-the-fly ab initio single-Hessian GWD to compute the photoelectron spectrum of the difluorocarbene anion and the absorption spectrum of methylamine.
In both cases we find that, despite the drastic approximations involved, the single Hessian GWD largely outperforms standard approaches for evaluating vibronic spectra based on global harmonic approximations. Whereas the global harmonic approaches fail miserably, the single Hessian GWD obtains a satisfactory agreement with the experiment. In addition, we identify which spectral features are sensitive to the choice of reference Hessian.
Conclusions are drawn in Sec.~\ref{sec:conclusion}.


\section{Single-Hessian Gaussian wavepacket dynamics \label{sec:theory}}

To accommodate readers with different interests, this section is divided into subsections focusing on different aspects of the single-Hessian GWD. In Secs.~\ref{subsec:spectroscopy} and \ref{subsec:GWD}, we outline the relationship between vibronic spectroscopy and Gaussian wavepacket dynamics. The single-Hessian approximation is reviewed in Secs.~\ref{subsec:SHA} and \ref{subsec:ref_hess}.
Together with Sec.~\ref{sec:integrators}, these sections target readers interested in the computational implementation and applications of single-Hessian GWD.
In contrast, Secs.~\ref{subsec:symplectic_structure} and \ref{subsec:geom_prop}, where we present symplectic and Hamiltonian perspectives on single-Hessian GWD, are directed toward readers interested in the theoretical foundations of the method.

\subsection{Time-dependent approach to vibronic spectroscopy\label{subsec:spectroscopy}}

The time-dependent approach to spectroscopy formulates the computation of spectra as a problem of quantum dynamics simulation, which requires solving the time-dependent Schr\"{o}dinger equation (TDSE) from the initial time up to an appropriate time that gives the desired spectral resolution.~\cite{Heller:1981a}
In the case of electronic absorption or photoelectron spectroscopies, the rotationally averaged absorption cross-section is evaluated as the Fourier transform~\cite{book_Tannor:2007}
\begin{equation}
    \sigma(\omega)=\frac{4\pi\omega}{3\hbar c}\mu^2\text{Re}\int_{0}^{\infty}C(t)\,e^{i(\omega+E_{i,0}/\hbar)t}\,dt 
    \label{eq:spectra}
\end{equation}
of the nuclear wavepacket autocorrelation function
\begin{equation}
    C(t)=\langle\psi_{0}|\psi_{t}\rangle,
    \label{eq:auto_corr}%
\end{equation}
where the wavepacket $\psi_{t}$ evolves under the TDSE
\begin{equation}
    i \hbar \dot{\psi}_{t}=\hat{H}\psi_{t}
    \label{eq:TDSE}
\end{equation}
with the final-state vibrational Hamiltonian $\hat{H}=T(\hat{p})+V(\hat{q})$. The kinetic term $T(p)=p^{T} \cdot m^{-1} \cdot p/2$ depends on the momentum $p$ and the real symmetric mass matrix $m$, and the Born-Oppenheimer potential energy surface $V(q)$ depends only on position $q$.
In Eq.~(\ref{eq:spectra}) we assumed the zero-temperature and Condon approximations, i.e., $E_{i,0}$ is the vibronic zero-point energy of state $\psi_{0}$ before photon absorption and $\mu=||\vec{\mu}_{fi}||$ is the magnitude of the electronic transition dipole moment evaluated at the initial-state equilibrium geometry. 


\subsection{Gaussian wavepacket dynamics\label{subsec:GWD}}

The evaluation of the autocorrelation function~(\ref{eq:auto_corr}) becomes straightforward when the time-dependent $D$-dimensional vibrational wavepacket is approximated by a Gaussian\cite{Heller:1975}
\begin{equation}
    \psi_{t}(q)= N_{t} \, \text{exp}\bigg[\frac{i}{\hbar}\bigg(\frac{1}{2}\,x^{T} \cdot P_{t} \cdot Q^{-1}_{t}\cdot x +p_{t}^{T} \cdot x+S_{t}\bigg)\bigg].
    \label{eq:HGWP}%
\end{equation}
Here, the Gaussian is expressed in position representation, using the ``$QPS$'' parametrization ($q_{t}$, $p_{t}$, $Q_{t}$, $P_{t}$, $S_{t}$);~\cite{Heller:1976a,Hagedorn:1980,book_Lubich:2008} $x:=q-q_{t}$ is the shifted position vector, $q_{t}$ and $p_{t}$ are the position and momentum  of the Gaussian's center, $Q_{t}$ and $P_{t}$ are two complex-valued $D$-dimensional matrices that determine the symmetric width matrix $A_{t}:=P_{t}\cdot Q_{t}^{-1}$ of the Gaussian, and $S_{t}$ is a real scalar. 
The complex prefactor $N_{t}=(\pi\hbar)^{-D/4} (\text{det} \, Q_{t})^{-1/2}$ results from an arbitrary but convenient gauge choice,~\cite{book_Lubich:2008,Lasser_Lubich:2020,Vanicek:2023} with the real part ensuring normalization of $\psi_{t}$, and the imaginary part contributing to the phase of the wavepacket.
Wavepacket~(\ref{eq:HGWP}) is the exact solution of the nonlinear TDSE~\cite{Vanicek:2023}
\begin{equation}
    i \hbar \dot{\psi}_{t}=\hat{H}_{\text{eff}}(\psi_{t})\psi_{t}
    \label{eq:nlTDSE}
\end{equation}
with a state-dependent Hamiltonian operator
\begin{equation}
    \hat{H}_{\text{eff}}(\psi_{t})=T(\hat{p})+V_{\text{eff}}(\hat{q};\psi_{t}),
    \label{eq:Heff}
\end{equation}
depending on the effective quadratic potential 
\begin{equation}
    V_{\text{eff}}(q;\psi_{t})=V_{0}+V_{1}^{T} \cdot x + x^{T} \cdot V_{2} \cdot x/2. \label{eq:Veff}
\end{equation}
The real scalar, vector, and symmetric-matrix coefficients $V_{0}$, $V_{1}$, and $V_{2}$ may depend on the state $\psi_{t}$ and are different for each method from the GWD family.~\cite{Vanicek:2023} 
They can be obtained by a general, two-step procedure: First, the exact potential $V$ is approximated with a state-dependent approximation $V_{\text{appr}}$. Then, the Dirac-Frenkel variational principle is applied to the Hamiltonian $\hat{H}_{\text{appr}}=T(\hat{p})+V_{\text{appr}}(\hat{q};\psi_{t})$ with the Gaussian ansatz (\ref{eq:HGWP}).
Different choices of $V_{\text{appr}}(q;\psi_{t})$ lead to different approximate methods, such as variational,~\cite{Coalson_Karplus:1990,Fereidani_Vanicek:2023} local cubic variational,~\cite{Pattanayak_Schieve:1994,Ohsawa_Leok:2013,Fereidani_Vanicek:2023a_v2} local harmonic,~\cite{Heller:1975,Kletnieks_Vanicek:2026} single-Hessian,~\cite{Begusic_Vanicek:2019,Vanicek:2023} or global harmonic GWD.
Here, we focus on the single-Hessian approximation of the potential to provide a firmer theoretical basis for the remarkable behavior exhibited by this method in practical applications. 
The global harmonic, local harmonic, and variational GWD, which will help us to validate the single-Hessian method in Sec.~\ref{sec:results}, are reviewed in Refs.~\onlinecite{Lasser_Lubich:2020} and ~\onlinecite{Vanicek:2023}.


\subsection{Single-Hessian GWD\label{subsec:SHA}}

In the local harmonic GWD, the evaluation of the local second derivative of the PES along the classical trajectory presents a major computational bottleneck. Therefore, global harmonic models remain the preferred choice for large molecules.
To increase the efficiency of the local harmonic GWD while still partially capturing anharmonicity, Begu\v{s}i\'{c}, Cordova, and Van\'i\v{c}ek introduced the single-Hessian approximation of the potential,~\cite{Begusic_Vanicek:2019}
\begin{align}
    V_{\text{SHA}}(q;q_{t}) & := V(q_{t})+V'(q_{t})^{T} \cdot x + x^{T} \cdot \kappa_{\text{ref}} \cdot x/2,
    \label{eq:V_SHA}
\end{align}
where the potential $V(q_{t})$ and gradient $V'(q_{t})$ are evaluated locally, but the Hessian is approximated with a constant matrix $\kappa_{\text{ref}}$.
Because the single-Hessian approximate potential is already quadratic, the application of the variational principle is redundant, hence $V_{\text{eff}}(q;\psi_{t})=V_{\text{appr}}(q;\psi_{t})$ for the single-Hessian approximation.\cite{Vanicek:2023}
Solving the TDSE~(\ref{eq:nlTDSE}) with the Gaussian~(\ref{eq:HGWP}) and the single-Hessian effective
potential~(\ref{eq:V_SHA}) yields the ordinary differential equations~\cite{Vanicek:2023}
\begin{align}
    \dot{q}_{t} & =m^{-1}\cdot p_{t},\label{eq:qEOM_SHA}\\
    \dot{p}_{t} & =-V^{\prime}(q_{t}),\label{eq:pEOM_SHA}\\
    \dot{Q}_{t}  &  =m^{-1}\cdot P_{t},\label{eq:QEOM_SHA}\\
    \dot{P}_{t}  &  =-\kappa_{\text{ref}}\cdot Q_{t},\label{eq:PEOM_SHA}\\
    \dot{S}_{t}  &  =T(p_{t})-V(q_{t}). \label{eq:SEOM_SHA}%
\end{align}
Similar to the local harmonic GWD, the Gaussian center $(q_{t},p_{t})$ evolves according to classical Hamilton's equations of motion with the true anharmonic potential $V(q_{t})$. Solving these equations requires a numerical propagation scheme.~\cite{Kletnieks_Vanicek:2026}
In contrast to the local harmonic GWD, the equations for the pair $(Q_{t},P_{t})$ are decoupled from the evolution of the center of the wavepacket and can be solved explicitly, as in the global harmonic GWD, resulting in quasi periodic oscillations of the width of the wavepacket around its initial value. Thus, in anharmonic systems, the single-Hessian GWD circumvents the nonphysical, unbounded growth of the width amplitude observed in the local harmonic GWD.~\cite{Ryabinkin_Genin:2024}
Typically, $\kappa_{\text{ref}}$ is replaced by $V^{\prime\prime}(q_\text{ref})$ at some reference position, which means that single-Hessian GWD requires only one Hessian evaluation for the entire propagation. Therefore, the computational cost of this method is comparable to that of classical molecular dynamics.

In its initial formulation,~\cite{Begusic_Vanicek:2019} single-Hessian GWD was introduced using Heller's original ``$A\gamma$'' parametrization ($q_{t}$, $p_{t}$, $A_{t}$, $\gamma_{t}$) for the wavepacket~\cite{Heller:1975}
\begin{equation}
    \psi_{t}(q)=\text{exp} \left[ \frac{i}{\hbar} \left( \frac{1}{2} x^{T}\cdot A_{t}\cdot x+p_{t}^T\cdot x+\gamma_{t} \right) \right],
    \label{eq:GWP}
\end{equation}
which employs a complex width matrix $A_{t}\equiv P_{t}\cdot Q_{t}^{-1}$, and combines the real phase $S_{t}$ and the complex prefactor $N_{t}$ in a single complex scalar $\gamma_{t}$.~\cite{book_Lubich:2008,Lasser_Lubich:2020,Vanicek:2023}
In this parametrization, Eqs.~(\ref{eq:QEOM_SHA})-(\ref{eq:SEOM_SHA}) are replaced by
\begin{align}
    \dot{A}_{t} & =-A_{t}\cdot m^{-1}\cdot A_{t}-\kappa_{\text{ref}},\label{eq:A_EOM_SHA}\\
    \dot{\gamma}_{t} & =T(p_{t})-V(q_{t})+(i\hbar/2)\,\text{Tr}(m^{-1}\cdot A_{t}).\label{eq:gamma_EOM_SHA}
\end{align}
The implementation in Turbomole~\cite{Begusic_Vanicek:2022} relies on the ``$QPS$'' parametrization and uses Eqs.~(\ref{eq:qEOM_SHA})-(\ref{eq:SEOM_SHA}).
The two parametrizations are equivalent under analytical propagation of the Gaussian width, but small differences in results may arise if numerical propagation schemes are used.~\cite{Fereidani_Vanicek:2023}
``$QPS$'' parametrization offers several advantages: (i)~the components of the width of the Gaussian evolve according to classical-like linear equations of motion [see Eqs.~(\ref{eq:QEOM_SHA}) and~(\ref{eq:PEOM_SHA})]. (ii)~Description of many properties of the Gaussian becomes simpler, because the matrices $Q_{t}$ and $P_{t}$ satisfy several remarkable mathematical relations.~\cite{book_Lubich:2008,Lasser_Lubich:2020}
(iii)~This parametrization forms the foundation of Hagedorn functions, which generalize Hermite functions to higher dimensions and form a complete orthonormal basis of Hilbert space $L^{2}(\mathbb{R}^{D})$.~\cite{Hagedorn:1998} 
In Appendix~\ref{sec:param}, we describe the equations of motion of single-Hessian GWD in a third alternative parametrization~\cite{Arickx_VanLeuven:1986} useful for one-dimensional systems.


\subsection{Reference Hessians} \label{subsec:ref_hess}

\begin{figure} 
\includegraphics{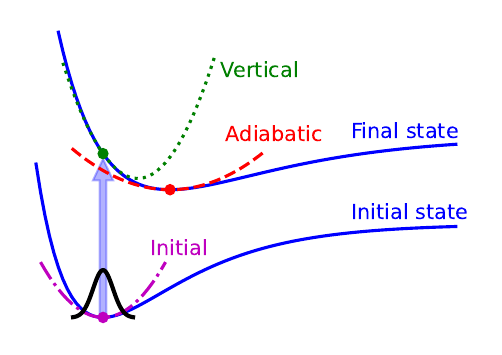}
\caption{Different choices of the reference ``single'' Hessian: the adiabatic Hessian (curvature of the red dashed curve) is evaluated at the minimum (red dot) of the final-state surface; the vertical Hessian (curvature of the green dotted curve) is evaluated at the Franck-Condon point (green dot); the initial Hessian (curvature of the magenta dash-dotted curve) is evaluated at the minimum (magenta dot) of the initial-state surface and is needed for constructing the initial wavefunction (black).}
\label{fig:Ref_geom}%
\end{figure}

Numerical applications of the single-Hessian method require selecting the reference ``single'' Hessian. 
Three natural options~\cite{Begusic_Vanicek:2019} are (see Fig.~\ref{fig:Ref_geom}) (i)~the adiabatic Hessian: Hessian $V_{f}^{\prime\prime}(q_{\text{eq},f})$ of the final PES evaluated at its minimum, (ii)~the vertical Hessian: Hessian $V_{f}^{\prime\prime}(q_{\text{eq},i})$ of the final PES evaluated at the Franck-Condon point, i.e., at the position of the minimum of the initial PES, and (iii)~the initial Hessian: Hessian $V_{i}^{\prime\prime}(q_{\text{eq},i})$ of the initial PES evaluated at its minimum.
These Hessians correspond to the curvature experienced by the system at three different key stages of the photodynamical process: before interaction with the electromagnetic field (initial Hessian), immediately after excitation (vertical Hessian), and after relaxation in the final electronic state (adiabatic Hessian).

The same Hessian choices have been widely used in conjunction with the global harmonic approximation.~\cite{AvilaFerrer_Santoro:2012,Baiardi_Barone:2013}
The ``adiabatic'' and ``vertical'' harmonic models are constructed using the homonymous Hessians, while several, different ``displaced'' harmonic oscillator models can be constructed using the initial Hessian, and differ only according to where the potential and gradient are evaluated.~\cite{Vanicek_Begusic:2021,Barone_Puzzarini:2021}

Like their harmonic counterparts, the adiabatic and vertical single-Hessian GWD are exact solutions in displaced, distorted, and Duschinsky-rotated harmonic oscillators.
The vertical single-Hessian GWD is attractive for practical reasons, as it avoids the need for final-state optimization, which can become challenging for electronically excited states.~\cite{AvilaFerrer_Santoro:2012,Wang_Durbeej:2020} Additionally, the vertical Hessian accurately captures the curvature of the PES at short times, making it well suited for modeling of the width dynamics in short-lived states. 
However, if the vertical Hessian contains imaginary frequencies, the Gaussian wavepacket spreads too rapidly, leading to a fast decay of the recurrences of the wavepacket autocorrelation function and to artificially broad peaks.~\cite{Begusic_Vanicek:2022}
The initial single-Hessian GWD is even more practical, because its Hessian is already required for constructing the initial wavepacket and because the Gaussian becomes frozen (i.e., its width does not change with time).~\cite{Vanicek:2023}
In contrast to adiabatic and vertical variants, the initial single-Hessian GWD is not an exact solution in general distorted and Duschinsky-rotated harmonic systems. 
However, whereas displaced global harmonic models completely neglect mode distortion and mixing between the two electronic states, an approximation that can significantly affect spectroscopy calculations,~\cite{Tapavicza:2019} the initial single-Hessian method captures the majority of these effects, as well as some anharmonicity, because it employs the exact anharmonic classical trajectory $(q_{t},p_{t})$ on the final-state surface.~\cite{Begusic_Vanicek:2019}

More generally, however, any Hessian matrix $\kappa_{\text{ref}}$ can be used to construct the single-Hessian approximate potential~(\ref{eq:V_SHA}). For example, one could employ a Hessian obtained at an electronic-structure level different from the level used to propagate the trajectory.


\subsection{Symplectic derivation of the equations of motion of single-Hessian GWD}
\label{subsec:symplectic_structure}

To show that the single-Hessian GWD equations of motion~(\ref{eq:qEOM_SHA})-(\ref{eq:SEOM_SHA}) are symplectic, we derive them as generalized Hamilton's equations of motion for the Hamiltonian function $h_{\text{SHA}}(z_{t}) := \langle\hat{H}_{\text{SHA}}\rangle_{\psi_{t}}$ on a symplectic manifold of Gaussian wavepackets equipped with a noncanonical symplectic structure. Once we show this, the symplecticity and conservation of effective energy follow automatically as every Hamiltonian vector field conserves both the symplectic structure and Hamiltonian function.~\cite{book_Cannas_Silva:2008}
We employ notation and procedure used by Ohsawa and Leok to obtain symplectic GWD starting from the local harmonic approximation.~\cite{Ohsawa_Leok:2013}
To simplify the equations in the derivation, in the following we omit the subscript $t$ on all parameters of the Gaussian.

In GWD, the symplectic manifold $M$ consists of unnormalized complex Gaussian wavepackets~(\ref{eq:GWP}) with parameters
\begin{equation}
    z:=(q,p,\mathcal{A},\mathcal{B},\phi,\delta)\label{eq:local_coord}
\end{equation} 
and squared norm $\mathcal{N}(\mathcal{B},\delta):=\lVert\psi\rVert^{2}=[\text{det}(\pi\hbar/\mathcal{B})]^{1/2}e^{-2\delta/\hbar}$. In Eq.~(\ref{eq:local_coord}), $q,p\in\mathbb{R}^{D}$, $\mathcal{A},\mathcal{B}\in\mathbb{R}^{D\times D}$ are real symmetric matrices such that $A = \mathcal{A}+i\mathcal{B}$, and $\phi,\delta\in\mathbb{R}$ are the real and imaginary parts of $\gamma$ (i.e., $\phi:=\text{Re}\,\gamma$ and $\delta:=\text{Im}\,\gamma$).~\cite{Ohsawa_Leok:2013}
The symplectic 2-form on $M$ is~\cite{Ohsawa_Leok:2013}
\begin{align}
    \omega(z)&:=\mathcal{N}(\mathcal{B},\delta)\left\{dq_{j}\wedge dp_{j}-p_{j}\,dq_{j}\wedge \text{Tr}(\mathcal{B}^{-1}d\mathcal{B})/2\nonumber\right.\\
    &~~~~ -(2/\hbar)p_{j}\,dq_{j}\wedge d\delta  \nonumber \\
    &~~~~ +(\hbar/8)(2\mathcal{B}_{jk}^{-1}\mathcal{B}_{lm}^{-1}+\mathcal{B}_{jm}^{-1}\mathcal{B}_{lk}^{-1})d\mathcal{A}_{jm}\wedge d\mathcal{B}_{kl} \nonumber \\
    &~~~~ +\left[\text{Tr}(\mathcal{B}^{-1}d\mathcal{A})\wedge d\delta-\text{Tr}(\mathcal{B}^{-1}d\mathcal{B})\wedge d\phi\right]/2\nonumber\\
    &~~~~ \left. +(2/\hbar)d\phi\wedge d\delta\right\},\label{eq:symp_form}
\end{align}
where we use Einstein's convention for sums over repeated indices.
The equations of motion for the parameters $q,p,\mathcal{A},\mathcal{B},\phi,\text{and}\,\delta$ are obtained from the definition $i_{X_{h_{\text{SHA}}}}\omega=dh_{\text{SHA}}\label{eq:H_Sys}$ of the Hamiltonian vector field $X_{h_{\text{SHA}}}$ on a Gaussian wavepacket manifold with the symplectic form~(\ref{eq:symp_form}) and Hamiltonian function
\begin{equation}
   h_{\text{SHA}}(z) := \langle\hat{H}_{\text{SHA}}\rangle_{\psi_{t}} =\langle \hat{T} \rangle + \langle \hat{V}_{\text{SHA}}\rangle = \mathcal{N}(\mathcal{B},\delta)\,\bar{h}_{\text{SHA}}(z),
   \label{eq:H_map} 
\end{equation}
where
\begin{align}
    \bar{h}_{\text{SHA}}(z)&=T(p)+V(q)+ (\hbar/4)\text{Tr}(\mathcal{A}\cdot m^{-1}\cdot \mathcal{A} \cdot \mathcal{B}^{-1} \nonumber\\
    &~~~  + \mathcal{B}\cdot m^{-1} + \kappa_{\text{ref}}\cdot \mathcal{B}^{-1}).\label{eq:H_map2}
\end{align}
Equating the explicit expressions for $i_{X_{h_{\text{SHA}}}}\omega$ and $dh_{\text{SHA}}$, which are provided in Appendix~\ref{sec:1-form}, yields Eqs.~(\ref{eq:qEOM_SHA}) and~(\ref{eq:pEOM_SHA}) for $\dot{q}$ and $\dot{p}$ as well as equations
\begin{align}
    \dot{\mathcal{A}}&=\mathcal{B}\cdot m^{-1}\cdot\mathcal{B}-\mathcal{A}\cdot
m^{-1}\cdot\mathcal{A}-\kappa_{\text{ref}},\\
\dot{\mathcal{B}}&=-(\mathcal{B}\cdot m^{-1}\cdot\mathcal{A}+\mathcal{A}\cdot m^{-1}\cdot\mathcal{B}),\\
\dot{\phi}&=T(p)-V(q)-(\hbar/2)\text{Tr}(m^{-1}\cdot\mathcal{B}),\\
\dot{\delta}&=(\hbar/2)\text{Tr}(m^{-1}\cdot\mathcal{A})
\end{align}
for $\dot{\mathcal{A}},\dot{\mathcal{B}},\dot{\phi},\text{and}\,\dot{\delta}$. 
Combining the equations of motion for $\mathcal{A}$ and $\mathcal{B}$ into a single equation for $A$ gives Eq.~(\ref{eq:A_EOM_SHA}), while doing the same for $\phi$ and $\delta$ gives the single equation~(\ref{eq:gamma_EOM_SHA}) for $\gamma$. Equations~(\ref{eq:A_EOM_SHA}) and~(\ref{eq:gamma_EOM_SHA}) (in ``$A\gamma$'' parametrization) can then be converted to Eqs.~(\ref{eq:QEOM_SHA})-(\ref{eq:SEOM_SHA}) in ``$QPS$'' parametrization.


\subsection{Geometric properties of the single-Hessian GWD} \label{subsec:geom_prop}

Regardless of the accuracy of an approximate solution, conservation of certain invariants (geometric properties) of the exact solution of the TDSE~(\ref{eq:TDSE}) is beneficial, as they govern the physical behavior of the system. 
The variational GWD,~\cite{Coalson_Karplus:1990,Lasser_Lubich:2020} which is the optimal approximate solution of the TDSE~(\ref{eq:TDSE}) with the Gaussian ansatz~(\ref{eq:HGWP}), is time-reversible and conserves the norm $\lVert\psi_{t}\rVert$, exact energy $E:=\langle\hat{H}\rangle_{\psi_{t}}$, effective energy $E_{\text{eff}}:=\langle\hat{H}_{\text{eff}}\rangle_{\psi_{t}}$, and the non-canonical symplectic structure $\omega$ [Eq.~(\ref{eq:symp_form})] of the finite-dimensional symplectic submanifold of Hilbert space formed by Gaussian wavepackets parametrized by Eq.~(\ref{eq:local_coord}).~\cite{Faou_Lubich:2006,Ohsawa_Leok:2013,Fereidani_Vanicek:2023}
From the definitions of the effective energy $E_{\text{eff}}$ and Hamiltonian function $h_{\text{SHA}}$ [Eq.~({\ref{eq:H_map}})] it is obvious that for the single Hessian approximation, the effective energy is the numerical value of the Hamiltonian function at the current coordinates of the Gaussian, i.e., $E_{\text{eff}}(t) = h_{\text{SHA}}(z_t)$. However, for clarity, we keep both symbols: $E_{\text{eff}}(t)$ is considered as a function of time, whereas $h_{\text{SHA}}(z)$ is considered as a function of coordinates. Another reason for keeping both symbols is that in the local harmonic GWD (as well as in other non-symplectic methods), one can still define the effective energy $E_{\text{eff}}$ as $\langle \hat{H}_{\text{LHA}}\rangle$, but there is no Hamiltonian function $h_{\text{LHA}}$ that would yield the equations of motion of the local harmonic GWD. (If one uses $h_{\text{LHA}}(z):=\langle \hat{H}_{\text{LHA}}\rangle$, one obtains the symplectic GWD of Ohsawa and Leok~\cite{Ohsawa_Leok:2013} rather than Heller’s local harmonic GWD.)

Let us discuss which of the properties of the variational GWD are also conserved by the single-Hessian and local harmonic GWD methods.
First, as any exact solution of the nonlinear TDSE~(\ref{eq:nlTDSE}), both the local harmonic and single-Hessian GWD are time-reversible and norm-conserving.~\cite{Vanicek:2023} 
As already shown in Sec.~\ref{subsec:symplectic_structure}, the equations of motion~(\ref{eq:qEOM_SHA})-(\ref{eq:SEOM_SHA}) of single-Hessian GWD can be derived as generalized Hamilton's equations for the Hamiltonian function (\ref{eq:H_map}) on the symplectic manifold of Gaussian wavepackets. Equation~(\ref{eq:H_map}) shows that this Hamiltonian function is given by the expectation value of the effective energy in a Gaussian with parameters $z_{t}$.
Because every Hamiltonian vector field on any symplectic manifold is a symplectic vector field,~\cite{book_Cannas_Silva:2008} single-Hessian GWD automatically conserves not only the effective energy $E_{\text{eff}}$ [given by the Hamiltonian function $h_{\text{SHA}}(z)$] but also the symplectic structure $\omega$.
However, single-Hessian GWD can also be derived differently by substituting the single-Hessian approximate potential~(\ref{eq:V_SHA}) and the Gaussian ansatz~(\ref{eq:HGWP}) into the TDSE~(\ref{eq:TDSE}). In this formulation, the conservation of $E_{\text{eff}}$ is not immediate and therefore requires explicit justification.
Such proof was provided in Ref.~\onlinecite{Begusic_Vanicek:2019} and also derived in Ref.~\onlinecite{Vanicek:2023} from the general expression
\begin{equation}
    \dot{E}_{\text{eff}}=\dot{V}_{0}-V_{1}^{T}\cdot\dot{q}_{t}+(\hbar/4)\,\text{Tr}(\dot{V}_{2}\cdot Q_{t}\cdot Q_{t}^{\dagger}), \label{eq:E_eff_conservation}
\end{equation}
for an arbitrary effective potential (\ref{eq:Veff}). In the single-Hessian approximation, $\dot{V}_{0}=V_{1}^{T}\cdot\dot{q}_{t}$, and $\dot{V}_{2}=d\kappa_{\text{ref}}/dt=0$ as $\kappa_{\text{ref}}$ is constant, hence $\dot{E}_{\text{eff}}=0$.
In contrast, the local harmonic GWD conserves neither $\omega$ nor $E_{\text{eff}}$.~\cite{Ohsawa_Leok:2013}

The time dependence of the \emph{exact} energy $E$ of a Gaussian wavepacket~(\ref{eq:HGWP}) is~\cite{Vanicek:2023}
\begin{equation}
    \dot{E}=\text{Re}\,\langle\hat{p}^{T}\cdot m^{-1}\cdot(\hat{V}^{\prime}-\hat{V}_{\text{eff}}^{\prime})\rangle.
    \label{eq:exactE}
\end{equation}
Neither the local harmonic, nor single-Hessian GWD conserves the exact energy in anharmonic potentials, because $V_{\text{SHA}}^{\prime}(q;q_{t})=V^{\prime}(q_{t}) +\kappa_{\text{ref}}\cdot (q-q_{t})\neq V^{\prime}(q)$.
However, unlike local harmonic GWD, single-Hessian GWD conserves the exact energy approximately in bounded regions of the potential. 
To justify this, we consider the exact Hamiltonian $\hat{H}$ to be a ``perturbed'' form of the effective Hamiltonian $\hat{H}_{\text{eff}}$. When the expected value of the perturbation
\begin{equation}
    E-E_{\text{eff}}:=\langle\hat{H}-\hat{H}_{\text{eff}}\rangle=\langle\hat{V}-\hat{V}_{\text{eff}}\rangle \label{eq:perturbation}
\end{equation}
remains small, the exact conservation of the effective energy $E_{\text{eff}}:=\langle\hat{H}_{\text{eff}}\rangle$ guarantees an approximate conservation of the energy $E:=\langle\hat{H}\rangle$.
Let us consider a local polynomial expansion $V(q)=\sum_{n=0}^{\infty}V^{(n)}(q_{t})_{k_{1}\dots k_{n}}\,x_{k_{1}}\dots x_{k_{n}}/n!$ of the potential, where Einstein's convention is used for sums over repeated indices $k_{i}\in({1,\dots,D})$. Within the single-Hessian approximation, the expected value~(\ref{eq:perturbation}) of the perturbation becomes
\begin{align}
    \langle\hat{V}-\hat{V}_{\text{SHA}}\rangle &= [V^{\prime\prime}(q_{t})-\kappa_{\text{ref}}]_{k_{1}k_{2}} \langle\hat{x}_{k_{1}}\hat{x}_{k_{2}}\rangle/2\nonumber\\
    &~~~+\sum_{n=2}^{\infty}V^{(2n)}(q_{t})_{k_{1}\dots k_{2n}}\langle \hat{x}_{k_{1}}\dots \hat{x}_{k_{2n}}\rangle/(2n)! \nonumber \\
    &= \text{Tr}\{[V^{\prime\prime}(q_{t})-\kappa_{\text{ref}}]\cdot\Sigma\}/2\nonumber\\
    &~~~ +\sum_{n=2}^{\infty}V^{(2n)}(q_{t})_{k_{1}\dots k_{2n}}\nonumber\\
    &~~~\times \Sigma_{k_{1}k_{2}}\dots\Sigma_{k_{2n-1}k_{2n}}/(2^{n}n!),
\end{align}
where $\Sigma=(\hbar/2)\,Q_{t}\cdot Q_{t}^{\dagger}$ denotes the position covariance matrix at time $t$. In the second step, we applied Isserlis's theorem~\cite{Isserlis:1918} to evaluate the multivariate Gaussian moments $\langle \hat{x}_{k_{1}}\dots \hat{x}_{k_{2n}}\rangle$.
When the wavepacket moves in a bounded region of the potential, the derivatives $V^{^{(2n)}}(q_{t})$ are oscillating, bounded functions.
Because the evolution of the width matrix of the single-Hessian Gaussian wavepacket exhibits only bounded quasi-periodic oscillations in time [see Eq.~(\ref{eq:mat_prod}) in Sec.~\ref{subsec:explicit_width} below], the position covariance matrix is also bounded. It follows that the expected value of the perturbation is bounded and that the exact energy of the single-Hessian GWD oscillates, but its oscillations are bounded. In other words, the energy does not drift and is approximately conserved over time.
The same argument does not hold for unbounded potentials, e.g., a cubic potential, where the entries of the tensors $V^{(2n)}(q_{\text{t}})$ may grow without limit over time.


\section{Geometric integrators for the single-Hessian GWD \label{sec:integrators}}

A peculiarity of single-Hessian GWD is the decoupling between the evolution of the center $(q_{t},p_{t})$ and width $(Q_{t},P_{t})$ of the wavepacket~(\ref{eq:HGWP}). As a consequence, the two pairs of variables may be propagated using independent schemes. We begin this section by presenting analytical expressions that yield the exact evolution of the width of a single-Hessian Gaussian wavepacket. These expressions are exact not only for single-Hessian but also for global harmonic GWD. Motivated by the extensive use of global harmonic models in spectroscopy,~\cite{AvilaFerrer_Santoro:2012,Baiardi_Barone:2013} we then also present an explicit, exact propagation scheme for $(q_{t},p_{t})$ and for the action $S_{t}$ in the harmonic potential. Finally, we discuss numerical time-stepping integration schemes that conserve the geometric properties of single-Hessian GWD in a general potential.


\subsection{Exact evolution of $Q_{t}$ and $P_{t}$ \label{subsec:explicit_width}}

Equations~(\ref{eq:QEOM_SHA}) and~(\ref{eq:PEOM_SHA}) can be combined into a single equation
\begin{equation}
\begin{pmatrix}
    \dot{Q}_{t} \\ \dot{P}_{t}
\end{pmatrix}
=
\begin{pmatrix}
0 & m^{-1}\\
-\kappa_{\text{ref}} & 0
\end{pmatrix}
\begin{pmatrix}
   Q_{t} \\ P_{t} 
\end{pmatrix},
\label{eq:width}
\end{equation}
which is an autonomous linear ordinary differential equation $\dot{Y}_{t}=K\cdot Y_{t}$ for the $2D\times D$ complex matrix $Y_{t}=(Q_{t}^{T},P_{t}^{T})^{T}$. This equation has solution $Y_{t}=\Phi_{t}(Y_{0})=\text{exp}(tK)\cdot Y_{0}=M(t)\cdot Y_{0}$.

The evolution matrix $M(t)$ is nothing else but the stability matrix of a harmonic system with Hessian $\kappa_{\text{ref}}$. $M(t)$ can be computed either by exponentiating matrix $tK$ directly, with numerically exact algorithms, or ``quasi-analytically'', as the matrix product~\cite{Tannor_Heller:1982}
\begin{align}
    M(t) & = \boldsymbol{m}\cdot\boldsymbol{L}\cdot\boldsymbol{\Omega}\cdot
    \begin{pmatrix}  \text{cos}(\Omega t) & \text{sin}(\Omega t) \\ -\text{sin}(\Omega t) & \text{cos}(\Omega t) \end{pmatrix} \nonumber\\
    &~~~ \cdot \boldsymbol{\Omega}^{-1}\cdot\boldsymbol{L}^{-1}\cdot\boldsymbol{m}^{-1},\label{eq:mat_prod}
\end{align}
where $\boldsymbol{m}=m^{-1/2}\oplus m^{1/2}$, $\boldsymbol{L}=L\oplus L$, $L$ is the transformation matrix that diagonalizes the mass-scaled reference ``single'' Hessian as
\begin{align}
    \Omega^2=L^{-1}\cdot m^{-1/2}\cdot \kappa_{\text{ref}}\cdot m^{-1/2}\cdot L,
\end{align}
$\Omega$ is the diagonal matrix containing the normal-mode frequencies, $\boldsymbol{\Omega}=I_{D}\oplus \Omega$, $I_{D}$ is the $D$-dimensional identity matrix, and $\oplus$ denotes the direct sum.

In the numerical examples presented in Sec.~\ref{sec:results}, we employ the numerical matrix exponential formulation [i.e., $M(t)=\text{exp}(tK)$], which is validated in Sec. S1 of the supplementary material; however, we have also verified the numerical equivalence of the two approaches (not shown).

\subsection{Exact evolution of $q_{t}$, $p_{t}$, and $S_{t}$ in a harmonic potential\label{subsec:explicit_center}}

In a global harmonic potential 
\begin{align}
    V_{\text{HA}}(q) &:=  V_{\text{ref}}+V_{\text{ref}}^{\prime T}%
    \cdot (q-q_{\text{ref}})\nonumber\\
    &~~~~+ (q-q_{\text{ref}})^{T} \cdot \kappa_{\text{ref}} \cdot (q-q_{\text{ref}})/2,\label{eq:V_GHA}
\end{align}
Eqs.~(\ref{eq:qEOM_SHA}) and~(\ref{eq:pEOM_SHA}) can also be solved exactly.
First, the potential~(\ref{eq:V_GHA}) is rewritten as
\begin{align}
    V_{\text{HA}}(q) & :=  V_{m}
    + (q-q_{m})^{T} \cdot \kappa_{\text{ref}} \cdot (q-q_{m})/2,\label{eq:V_AHA}
\end{align}
where $V_{m}:=V_{\text{ref}}-V_{\text{ref}}^{\prime T}\cdot\kappa_{\text{ref}}^{-1}\cdot V_{\text{ref}}^{\prime}/2$ is the potential energy at the minimum, which is located at $q_{m}:=q_{\text{ref}}-\kappa_{\text{ref}}^{-1}\cdot V_{\text{ref}}^{\prime}$.
Then, the center of the wavepacket evolves according to~\cite{Tannor_Heller:1982}
\begin{align}
    \begin{pmatrix}  q_{t}-q_{m} \\ p_{t} \end{pmatrix} & = M(t)\cdot
    \begin{pmatrix}  q_{0}-q_{m} \\ p_{0} \end{pmatrix}.
\end{align}
In the harmonic potential~(\ref{eq:V_AHA}), the right-hand side of Eq.~(\ref{eq:SEOM_SHA}) can be developed as
\begin{align}
    T(p_{t})-V_{\text{HA}}(q_{t}) &= [p_{t}^{T}\cdot\dot{q}_{t} - V'_{\text{HA}}(q_{t})^{T}\cdot (q_{t}-q_{m})]/2 - V_{m} \nonumber\\
    &= (1/2)\,d[p_{t}^{T}\cdot(q_{t}-q_{m})]/dt-V_{m}.
\end{align}
Thus, Eq.~(\ref{eq:SEOM_SHA}) has an analytical solution
\begin{equation}
    S_{t} = S_{0} + \left[p_{t}^{T}\cdot(q_{t}-q_{m})-p_{0}^{T}\cdot(q_{0}-q_{m})\right]/2-t\,V_{m}.
\end{equation}


\subsection{Geometric numerical integrators\label{subsec:geom_integrators}}

In a general, anharmonic potential $V(q)$, equations of motion for $q_{t}$, $p_{t}$, and $S_{t}$ must be solved numerically. When equations~(\ref{eq:qEOM_SHA})-(\ref{eq:SEOM_SHA}) are solved numerically, conservation of the geometric properties of the single-Hessian GWD is not guaranteed.~\cite{book_Lubich:2008}

Reference~\onlinecite{Vanicek:2023} described geometric integrators of arbitrary order of accuracy for the nonlinear TDSE~(\ref{eq:nlTDSE}) with a general effective potential~(\ref{eq:Veff}) based on a standard combination of splitting and composition techniques.~\cite{book_Hairer_Wanner:2006,book_Iserles:2009,book_Sanz-Serna_Calvo:2018,book_Blanes_Casas:2016} These integrators directly generalize the second-order splitting integrators proposed earlier for the variational GWD;~\cite{book_Lubich:2008,Faou_Lubich:2006} in addition, Ref.~\onlinecite{Vanicek:2023} described the conditions on the general effective potential~(\ref{eq:Veff}) for which these integrators are explicit.
The only assumption was that the coefficients $V_{0}$, $V_{1}$, and $V_{2}$ depend on the wavepacket $\psi_{t}$ only through the parameters $q_{t}$ and $Q_{t}$ of the Gaussian; this condition is satisfied by the single-Hessian effective potential~(\ref{eq:V_SHA}), which depends only on $q_{t}$.
The geometric integrators are obtained by decomposing the Hamiltonian flow, which, in general, cannot be solved explicitly, into a sequence of explicitly solvable kinetic and potential flows. 
For the single-Hessian GWD, the flow associated with potential propagation, during which $\hat{H}_{\text{eff}}=V_{\text{SHA}}(\hat{q};q_{t})$, is
\begin{align}
    q_{t}  &  =q_{0},\label{eq:q_EOM_V}\\
    p_{t}  &  =p_{0}-t\,V^{\prime}(q_{0}),\\
    Q_{t}  &  =Q_{0},\\
    P_{t}  &  =P_{0}-t\,\kappa_{\text{ref}}\cdot Q_{0},\label{eq:P_EOM_V}\\
    S_{t}  &  =S_{0}-t\,V(q_{0}).
\end{align}
The flow for kinetic propagation, during which $\hat{H}_{\text{eff}}=T(\hat{p})$, is given by Eqs. (132), (133), (163)-(165) from Ref.~\onlinecite{Vanicek:2023}:
\begin{align}
    q_{t} & =q_{0}+tm^{-1}\cdot p_{0},\\
    p_{t} & =p_{0},\\
    Q_{t} & =Q_{0}+tm^{-1}\cdot P_{0},\\
    P_{t} & =P_{0},\\
    S_{t}  &  =S_{0}+t\,T(p_{0}).
\end{align}
Being exact solution of the (non)linear TDSE~(\ref{eq:nlTDSE}) with its associated effective Hamiltonian, each flow individually preserves all geometric properties of the single-Hessian method. 

The simplest such algorithms are the first-order TV and VT algorithms, constructed by composing the kinetic (T) and potential (V) propagations with the same time step $\Delta t$. Since they are not symmetric, these integrators are not time-reversible.
Composing the VT and TV algorithms, each with time step $\Delta t/2$, yields the symmetric, second-order TVT and VTV algorithms, which are generalizations of the St\"{o}rmer-Verlet algorithm of classical Hamiltonian dynamics.\cite{book_Lubich:2008} 
Well-chosen symmetric compositions of the second-order algorithms then yield integrators of higher, even orders of accuracy.~\cite{Yoshida:1990,Suzuki:1990,Choi_Vanicek:2019,Roulet_Vanicek:2019}
Due to the splitting, $H_{\text{eff}}$ is explicitly time-dependent, and the integrators conserve the effective energy only approximately, with an $\mathcal{O}(\Delta t^{M})$ error, where the power $M$ is greater than or equal to the order of the algorithm.~\cite{book_Hairer_Wanner:2006,Roulet_Vanicek:2019,Choi_Vanicek:2019}
All other geometric properties of the single-Hessian GWD are preserved exactly by the symmetric integrators.~\cite{Vanicek:2023} 

Fourth- and higher-order integrators can be constructed via different recursive schemes,~\cite{Yoshida:1990,Suzuki:1990} though the number of splitting substeps grows exponentially with the order of convergence. Therefore, we use more efficient, non-recursive composition schemes~\cite{Kahan_Li:1997,Sofroniou_Spaletta:2005,Choi_Vanicek:2019} for integrators of order higher than four. 

If one is interested in computing observables at a specific time $t$, the best approach is to use a combination of numerical stepwise propagation of the center of the wavepacket with exact global evolution of the width matrix. As shown in Fig.~S1 of the supplementary material, this hybrid approach agrees with the fully converged numerical propagation, but the hybrid approach is much more efficient because it can obtain the final width matrix in one step.
In contrast, the fully numerical approach remains nearly as efficient as the hybrid approach in spectroscopy calculations, because the spectrum~(\ref{eq:spectra}) depends on the values of the autocorrelation function~(\ref{eq:auto_corr}) at all times and therefore even the width must be evaluated frequently, i.e., at each time step $\Delta t$, whose size is determined by the range of the spectrum.


\subsection{Conservation of the symplectic structure by the geometric integrators\label{subsec:num_symplectic_structure}} 

For a geometric integrator, $z_{t}=\Phi_{t}(z_{0})$ denotes the composed flow consisting of many steps, each of which, in turn, is a composition of several kinetic and potential substeps, i.e., $\Phi = \Phi_{n}\circ\dots\circ\Phi_{1}$.
The conservation of the symplectic structure by the kinetic and potential flows follows automatically from the Hamiltonian construction (Sec.~\ref{subsec:symplectic_structure}) of the single-Hessian method with Hamiltonians $\langle\hat{T}\rangle$ and $\langle\hat{V}_{\text{SHA}}\rangle$, respectively. If both kinetic and potential flows are symplectic, then any composition $\Phi$  of them is symplectic.

While this general discussion already proves the conservation of symplectic structure by the geometric integrators developed for the single-Hessian GWD, an alternative, explicit proof provides a convenient starting point for numerical verification of the conservation of the symplectic structure by a specific integrator.
For simplicity, we omit the phase factor $\exp(iS_{t}/\hbar)$ in (\ref{eq:HGWP}) and work in an equivalent representation of the Gaussian wavepacket (without a phase) in a projective Hilbert space with a phase symmetry.~\cite{Ohsawa:2015,Ohsawa:2015a}
Without the parameter $S_t$, the symplectic manifold parametrized by the set $z_{t}=(q_{t},p_{t},Q_{t},P_{t})$ has the simpler, reduced symplectic form~\cite{Ohsawa:2015a}
\begin{align}
    \omega(z_{t})  &  =dq_{j}\wedge dp_{j}+(\hbar/2)\,d\text{Re}Q_{jk}\wedge d\text{Re}P_{jk}\nonumber\\
    &~~~ +(\hbar/2)\,d\text{Im}Q_{jk}\wedge d\text{Im}P_{jk}, 
    \label{eq:sympl_QPS}%
\end{align}
where we again use Einstein's convention for sums over repeated indices and omit the subscript $t$ on all parameters for simplicity.
An integrator with flow $z_{t}=\Phi_{t}(z_{0})$ is symplectic if the Jacobian $\Phi_{t}^{\prime}(z_{0})$ of the flow satisfies the condition~\cite{book_Hairer_Wanner:2006}
\begin{equation}
\Phi_{t}^{\prime}(z_{0})^{T}\cdot\omega(z_{t})\cdot\Phi_{t}^{\prime}%
(z_{0})=\omega(z_{0}). \label{eq:JBJ=B}%
\end{equation}
For a geometric integrator based on a composition of kinetic and potential propagations, the Jacobian $\Phi_{t}^{\prime}(z_{0})$ is obtained by matrix multiplication of the Jacobians of all kinetic and potential substeps, i.e., $\Phi^{\prime}=\Phi_{n}^{\prime}\cdot\dots\cdot\Phi_{1}^{\prime}$.
For the potential flow [Eqs.~(\ref{eq:q_EOM_V})-(\ref{eq:P_EOM_V})], the Jacobian is the $(2D+4D^{2})$-dimensional matrix
\begin{equation}
    \Phi_{\text{V}_{\text{SHA}},t}^{\prime}(z_{0})=I_{2D+4D^{2}}-t\left[
    \begin{pmatrix}
    0 & 0 \\
    V^{\prime\prime}(q_{0}) & 0
    \end{pmatrix}\oplus b \oplus b\right]
    , \label{eq:HJV}%
\end{equation}
where \begin{equation}
    b=%
    \begin{pmatrix}
    0 & 0\\
    \kappa_{\text{ref}} \otimes I_{D} & 0%
    \end{pmatrix}
    \label{eq:d_subblock}%
\end{equation}
and $\otimes$ denotes the Kronecker product. 
Since the Hessian matrices are symmetric, Eq.~(\ref{eq:JBJ=B}) holds for the Jacobian~(\ref{eq:HJV}), and the potential flow of the single-Hessian GWD is symplectic.
Analogous proof of the conservation of the symplectic structure by the kinetic flow, with Jacobian
\begin{equation}
    \Phi_{\text{T},t}^{\prime}(z_{0})= M_{2D} \oplus M_{2D^{2}} \oplus M_{2D^{2}}, \label{eq:HJT}%
\end{equation}
where
\begin{equation}
    M_{2D}=%
    \begin{pmatrix}
    I_{D} & t\,m^{-1}\\
    0 & I_{D}%
    \end{pmatrix}
    \label{eq:Stability}%
\end{equation}
is the stability matrix and $M_{2D^{2}}=M_{2D}\otimes I_{D}$, can be found in Ref.~\onlinecite{Fereidani_Vanicek:2023a_v2}. 

Although this ``explicit'' proof was not strictly necessary, it provides the tools for testing symplecticity of numerical simulations. To verify conservation by a numerical flow $\Phi_{t}(z_{0})$, obtained by composing many kinetic and potential steps, one simply computes the distance
\begin{equation}
    \lVert\Phi_{t}^{\prime}(z_{0})^{T}\cdot\omega(z_{t})\cdot\Phi_{t}^{\prime}(z_{0})-\omega(z_{0})\rVert \label{eq:Frob_distance},
\end{equation}
which measures the accuracy with which Eq.~(\ref{eq:JBJ=B}) is satisfied. We refer to this distance as ``symplecticity''.~\cite{Fereidani_Vanicek:2023,Fereidani_Vanicek:2023a_v2}


\section{Computational details \label{sec:details}}

Simulations were performed using our in-house Fortran 2018 code for molecular quantum dynamics.

\subsection{Morse potential}

Before moving to ab initio simulations, we investigated the Morse potential
\begin{equation}
V(q)=\frac{\omega_{e}}{4\chi}\big\{1-\exp\big[-(2m\omega_{e}\chi)^{1/2}(q-q_{\text{eq}})\big]\big\}^{2},\label{eq:1D_Morse}%
\end{equation}
with a minimum at the equilibrium position $q_{\text{eq}}=0$ and depending on the harmonic frequency $\omega_{e}=0.9$ and dimensionless anharmonicity $\chi=0.02$. 
All values are presented in natural units (n.u.) with $m = 1$.
The initial wavepacket was a Gaussian~(\ref{eq:HGWP}) with $q_{0}=10$, $p_{0}=0$, $Q_{0}=1$, and $P_{0}=i$, i.e., the eigenstate of a harmonic potential with frequency $\omega_{g}=1$, and we analyzed the accuracy of the single-Hessian GWD for different values of the semiclassical parameter $\hbar$. The wavepacket was propagated for a total time of 100 (i.e., for $100000$ steps with a time step $\Delta t=0.001$) in the adiabatic single-Hessian, local harmonic, and variational effective potentials, using the second-order TVT integrator described in Sec.~\ref{subsec:geom_integrators}, or with quantum dynamics using the analogous second-order split operator algorithm.~\cite{Kosloff_Kosloff:1983} The position grid for quantum dynamics consisted of $512/\hbar$ equidistant points between $-12$ and $63$. 
The momentum grid contained the same number of points, equally spaced between $-21.4$ and $21.4$. These parameters were chosen to ensure adequate convergence, starting from the classical turning points located at $-3.2$ and $10$ in position space and at $\pm4.0$ in momentum space. 

\subsection{On-the-fly ab initio calculations}

Performing on-the-fly ab initio Gaussian wavepacket dynamics requires an interface between an electronic structure program, which evaluates the PES information, and our in-house code, which reads this information, transforms between Cartesian and normal-mode coordinates, and performs the dynamics. The implementation has been detailed in Refs.~\onlinecite{Wehrle_Vanicek:2014,Wehrle_Vanicek:2015}.
For all calculations, the reference Eckart frame was the equilibrium geometry on the PES of the electronic initial state.

For ammonia, the ground-state geometry was optimized using the density functional theory, while the time-dependent density functional theory was used for calculations in the first excited electronic state; the functional was CAM-B3LYP (Ref.~\onlinecite{Yanai_Handy:2004}) and the basis set was 6-31+G**,~\cite{Frisch_Binkley:1984} as implemented in Gaussian 16 package.\cite{Frisch_Fox:2016}
This system was already studied in Ref.~\onlinecite{Begusic_Vanicek:2019}, where the authors computed the absorption spectrum from the ground to the excited state at the CASPT2(8e,8o) level of theory using the augmented correlation-consistent polarized valence triple-zeta (aug-cc-pVTZ) basis set.
Rather than obtaining highly accurate spectra of ammonia, as in Ref.~\onlinecite{Begusic_Vanicek:2019}, our goal here was to perform a systematic analysis of the geometric integrators for several orders of composition and for several time steps. To this end, we selected a level of theory that is computationally efficient while still providing a qualitatively correct PES.
In contrast, for the other molecules, whose spectra are computed with the single-Hessian GWD for the first time, the electronic structure method was chosen to yield quantitative results.
For difluorocarbene, all electronic structure calculations were performed with Gaussian 16 using a hybrid Perdew-Burke-Ernzerhof-type (PBE0) functional~\cite{Adamo_Barone:1999} and the aug-cc-pVTZ basis set,~\cite{Kendall_Harrison:1992} as in Ref.~\onlinecite{Zhang_Vanicek:2025}.
The ab initio electronic structure calculations for methylamine were performed at the multistate (MS-)CASPT2(4e,6o) level~\cite{Finley_Andres:1998} with 6-31++G** basis set,~\cite{Frisch_Binkley:1984} as implemented in Molpro2019 package;~\cite{MOLPRO:2019_v2} a level shift of 0.3 a.u. was applied. The state averaging was performed for the reference CASSCF calculation by setting equal weight for the first two singlet states.~\cite{Kletnieks_Vanicek:2026a} 
See the supplementary material for the optimized geometries of all species and for the harmonic vibrational frequencies of difluorocarbene and methylamine.

The initial wavepacket was always chosen as the vibrational ground state of the harmonic fit at the minimum of the PES of the electronic initial state.
The propagation times were $4096 \, \textrm{a.u.} \approx 99 \, \textrm{fs}$ for ammonia, $16000 \, \textrm{a.u.} \approx 387 \, \textrm{fs}$ for difluorocarbene, and $10000 \, \textrm{a.u.} \approx 242 \, \textrm{fs}$ for methylamine.
Single-Hessian and local harmonic GWD were performed using the numerical integrators from  Sec.~\ref{subsec:geom_integrators}, while global harmonic GWD was performed using the exact propagation scheme described in Secs.~\ref{subsec:explicit_width} and~\ref{subsec:explicit_center}.

To obtain smooth spectra in the frequency domain, the correlation functions were padded with zeros up to $80000\, \textrm{a.u.}$
The computed spectra were broadened using a Lorentzian with a half width at half maximum of $76\, \textrm{cm}^{-1}$ for difluorocarbene and a Gaussian with a half width at half maximum of $55\, \textrm{cm}^{-1}$ for methylamine.
To account for the error in the ab initio electronic structure estimate of the vertical excitation energies, the adiabatic harmonic spectra were shifted to match the experimental origin, while all other computed spectra were shifted to match the most intense experimental peak using a procedure from Ref.~\onlinecite{Zhang_Vanicek:2025}; the shifts are listed in Tables S5 and S9 of the supplementary material.


\section{Numerical examples \label{sec:results}}

\subsection{Absorption spectra of diatomic molecules\label{subsec:PT}}

\begin{figure*} 
\includegraphics{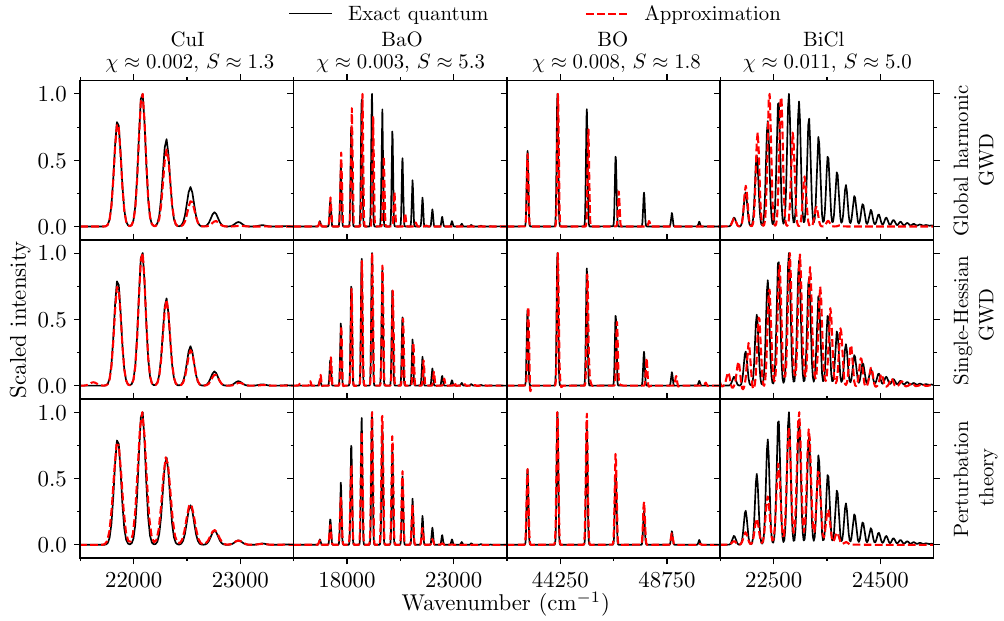}
\caption{Vibronic absorption spectra of CuI, BaO, BO, and BiCl obtained from numerically exact quantum calculations, global harmonic GWD, single-Hessian GWD, and second-order vibrational perturbation theory. For each molecule, the anharmonicity parameter $\chi$ and Huang-Rhys displacement parameter $S$ are reported.}
\label{fig:Morse_PT}%
\end{figure*}

Despite being a crude semiclassical approximation, single-Hessian GWD has proven sufficiently accurate for calculating low- to medium-resolution vibronic spectra of many weakly anharmonic, polyatomic molecules.~\cite{Begusic_Vanicek:2019,Prlj_Vanicek:2020,Begusic_Vanicek:2021}
Here, we study spectra of four diatomic molecules. Showing only one vibrational progression, these spectra are inherently fully vibrationally resolved and therefore constitute an almost pathological test case for single-Hessian GWD.
Figure~\ref{fig:Morse_PT} compares spectra of CuI, BaO, BO, and BiCl computed using single-Hessian GWD with spectra obtained from numerically exact quantum calculations, global harmonic GWD, and second-order vibrational perturbation theory,~\cite{Barone:2004} a state-of-the-art approach for incorporating anharmonicity in spectroscopy simulations. 
For each system, the ground-state PES was approximated by a harmonic potential, while the excited-state PES was modeled using a Morse potential~(\ref{eq:1D_Morse}), with parameters taken from Ref.~\onlinecite{Huber_Herzberg:1979} (see Table S1 of the supplementary material).
These molecules were selected to assess the accuracy of the approximate methods in regions of the PES where anharmonic effects become significant. In vibronic spectroscopy, anharmonicity is commonly characterized by the dimensionless anharmonicity parameter $\chi$ and the Huang-Rhys factor $S=m\omega_{g}(q_{\text{eq}}-q_{0})^{2}/(2\hbar)$, another dimensionless parameter characterizing the displacement between the minima of the ground- and excited-state surfaces. 
Because eigenenergies of a Morse potential are $E_{n}=\hbar\omega(n+1/2)-\hbar\omega\chi(n+1/2)^{2}$, increasing $\chi$ leads to a more rapid decrease in peak spacing, while increasing $S$ results in a wider spectral range, by allowing transitions to higher-excited, more anharmonic vibrational states.
Global harmonic models suffer from two fundamental limitations. First, these models predict a constant spacing between peaks, entirely independent of the anharmonicity parameter $\chi$. Second, they yield an incorrect spectral envelope, since the wavepacket evolves on a harmonic surface, an approximation which becomes increasingly inaccurate as the displacement $S$ grows.
While both perturbation theory and single-Hessian GWD always outperform global harmonic calculations, these methods achieve different accuracy in the four applications.
Perturbation theory correctly captures the decreasing peak spacing. However, due to its time-independent formulation, this method does not capture the correct spectral envelope at large displacements. Relying on third and fourth derivatives of the PES, perturbation theory can only access a limited number of vibrational excitations, extending to at most four additional states beyond the harmonic approximation.
Single-Hessian GWD does not reproduce the decreasing peak spacing. Instead, this method predicts an effective constant spacing that differs from the harmonic value. This spacing corresponds to an average anharmonic frequency, determined by the period of the wavepacket motion on the true anharmonic PES, and therefore depends on both $\chi$ and $S$. However, single-Hessian GWD accurately captures the spectral envelope, as the wavepacket follows the exact classical trajectory on the anharmonic surface.
These observations highlight the complementary strengths of perturbation theory and single-Hessian GWD: the former incorporates anharmonicity in a static manner, while the latter accounts for anharmonicity dynamically.

As stated previously, single-Hessian GWD is better suited for computing spectra of large polyatomic molecules. Such spectra exhibit a vast number of overlapping vibrational progressions that are not resolved experimentally due to broadening effects, and therefore full vibrational resolution is not needed.
Although the limitations discussed above remain valid, further practical considerations should be taken into account. Perturbation theory requires the explicit computation of all individual spectral lines, many of which will ultimately be indistinguishable after broadening. Such procedure involves unnecessary computational effort. In contrast, single-Hessian GWD inherently incorporates a degree of spectral broadening through its dynamical formulation. As a consequence, precise knowledge of individual spectral lines is not required, and the effective anharmonic spacing provided by this method is often sufficient.

\subsection{Geometric properties and semiclassical errors of single-Hessian GWD \label{subsec:Morse}}

\begin{figure} 
\includegraphics{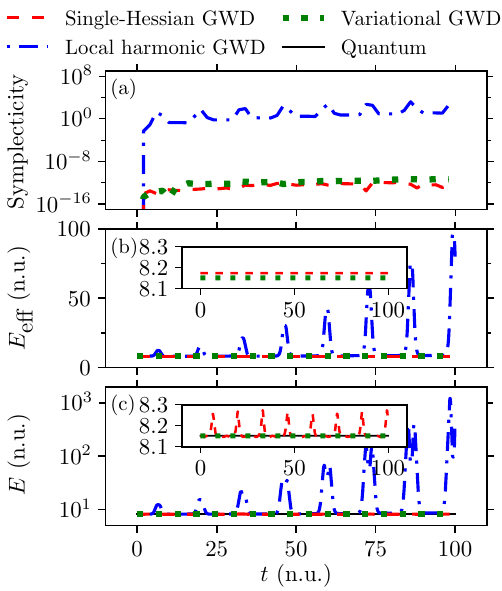}
\caption{Conservation of (a) symplectic structure, (b) effective energy, and (c) energy of a Gaussian wavepacket propagated with different versions of GWD in a Morse potential~(\ref{eq:1D_Morse}), for $\hbar=0.1$. Panel (c) shows the conservation of energy also for a Gaussian wavepacket propagated exactly quantum-mechanically.}
\label{fig:Morse_geom_prop}%
\end{figure}

The unexpected accuracy of the single-Hessian approach in spectroscopy simulations has been attributed to its superior geometric properties over the local harmonic version.~\cite{Vanicek:2023}
To assess the performance of the two methods, we considered dynamics in the Morse potential~(\ref{eq:1D_Morse}). The initial conditions were chosen such that the curvature of the potential changes sign during the dynamics. This choice provides a stringent test for single-Hessian GWD, because thawed GWD is known to deteriorate in such situations.~\cite{Begusic_Vanicek:2022}
Panels (a) and (b) of Fig.~\ref{fig:Morse_geom_prop} confirm numerically that the single-Hessian GWD conserves both the effective energy and the symplectic structure. 
Thus, despite the lower computational cost compared to the local harmonic variant, the behavior of single-Hessian GWD more closely resembles that of the optimal (and much more expensive) variational GWD. 
While the single-Hessian GWD conserves exactly the symplectic structure $\omega$ and effective energy $E_{\text{eff}}$, the energy $E$ is known to be conserved exactly only by the variational GWD. It is therefore reassuring that, whereas the local harmonic GWD displays an uncontrollable increase in energy, the single-Hessian GWD nearly conserves the energy over time [see panel (c)].
Note that the energy of a local harmonic Gaussian wavepacket diverges from its effective energy. 
As explained in Sec.~\ref{subsec:geom_prop}, this behavior is due to the unbounded dynamics of the width of the wavepacket.~\cite{Ryabinkin_Genin:2024}

\begin{figure} 
\includegraphics{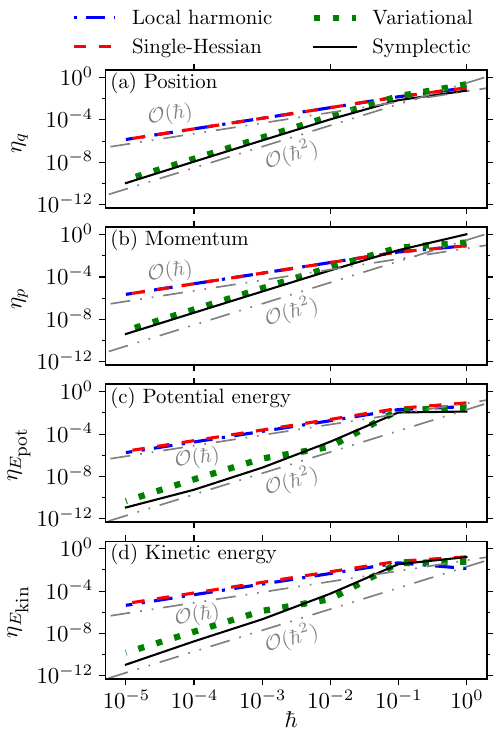}
\caption{Convergence of expectation values of (a) position $q$, (b) momentum $p$, (c) potential energy $E_{\text{pot}}$, and (d) kinetic energy $E_{\text{kin}}$ with respect to the semiclassical parameter $\hbar$ for a Gaussian wavepacket propagated in a Morse potential~(\ref{eq:1D_Morse}) with different GWD methods [single-Hessian,~\cite{Begusic_Vanicek:2019} local harmonic,~\cite{Heller:1975} variational,~\cite{Coalson_Karplus:1990} and symplectic semiclassical~\cite{Ohsawa_Leok:2013} (also known as extended semiclassical~\cite{Pattanayak_Schieve:1994} or local cubic variational~\cite{Fereidani_Vanicek:2023a_v2}) GWD]. Convergence is measured by the error $\eta_{A}=|(A_{\text{SC}}-A_{\text{QM}})/A_{\text{QM}}|$, where $A_{\text{SC}}$ denotes the expectation value of the observable $A$ at time $t = 10$ obtained using the semiclassical GWD method and $A_{\text{QM}}$ denotes the exact quantum expectation value.}
\label{fig:Morse_hbar}%
\end{figure}

Although both local harmonic and variational GWD achieve $\mathcal{O}(\hbar^{1/2})$ error for the wavefunction,~\cite{Hagedorn:1998,book_Lubich:2008} variational GWD is more accurate for evaluating the observables,~\cite{Ohsawa:2021,Burkhard_Lasser:2024} attaining $\mathcal{O}(\hbar^{2})$ error~\cite{Burkhard_Lasser:2024} compared to the $\mathcal{O}(\hbar)$ error of local harmonic GWD. 
The second-order convergence was proven analytically in Ref.~\onlinecite{Burkhard_Lasser:2024} (the order was originally established as $m\geq3/2$ for symplectic GWD~\cite{Ohsawa:2021} and subsequently refined to $m=2$ for variational GWD~\cite{Burkhard_Lasser:2024}). However, previous numerical studies of this exponent were not decisive,~\cite{King_Ohsawa:2020} because for $m=2$ or $3/2$ they require extremely accurate quantum benchmarks calculations. Therefore, in Fig.~\ref{fig:Morse_hbar} we confirm numerically this remarkable theoretical result for the variational GWD and show that symplectic GWD also has quadratic convergence--not only for energy~\cite{Ohsawa:2021} but also for position and momentum. To obtain quantum results with accuracy near machine precision, we exploited the exponential convergence of the dynamical Fourier method with respect to the grid density.
Given that the single-Hessian method is computationally cheaper than the local harmonic variant, one might expect that the gain in efficiency comes at the expense of accuracy. However, our numerical results for position, momentum, and potential and kinetic energies in Fig.~\ref{fig:Morse_hbar} suggest that the $\mathcal{O}(\hbar)$ accuracy for observables is preserved. This is another pleasing property of single-Hessian GWD. 

\subsection{Application of geometric integrators in on-the-fly ab initio simulations \label{subsec:NH3}}

Single-Hessian GWD makes it possible to perform semiclassical simulations at the cost of classical ab initio molecular dynamics. To determine whether the properties of geometric integrators designed for the single-Hessian GWD survive in an ab initio setting, we performed on-the-fly ab initio dynamics in the first excited state of ammonia using either the symplectic integrators from Sec.~\ref{subsec:geom_integrators} or the non-symplectic, yet popular fourth-order Runge-Kutta method.
Ammonia was selected because it was the first molecule on which ab initio single-Hessian GWD was validated;~\cite{Begusic_Vanicek:2019} therefore, we could safely investigate the geometric properties without concerns about limitations of the method.
Because the qualitative analysis of the properties of the integrators should not be affected by the accuracy of the electronic structure, we opted for an economical combination of theory and basis set (see Sec.~\ref{sec:details}), which permitted us to perform a thorough analysis for several integrators and time steps.
Although the kinetic and potential flows from Sec.~\ref{subsec:geom_integrators} can be composed to obtain integrators of arbitrary order of accuracy, to avoid cluttered plots we only show the results of the second-order VTV algorithm and of its fourth- and eighth-order compositions.
In all calculations, the single-Hessian potential was constructed using the adiabatic Hessian.

\begin{figure} 
\includegraphics{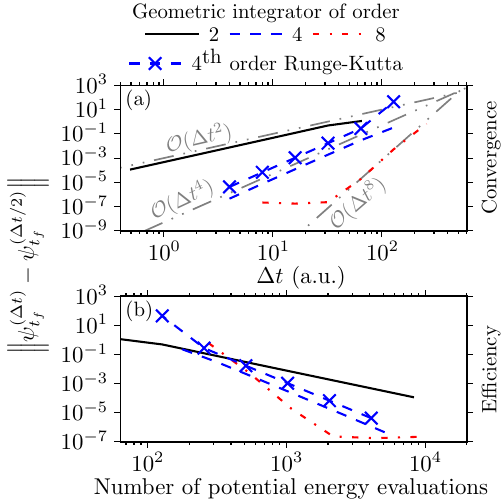}
\caption{Convergence and efficiency of the geometric integrators and of the fourth-order Runge-Kutta method for the adiabatic single-Hessian GWD in the first excited state of ammonia. Convergence [panel (a)] is measured by the convergence error~(\ref{eq:conv_err}) at the final time $t_{f} =4096 \, \text{a.u.}$ as a function of the time step $\Delta t$. 
Efficiency [panel (b)] is measured by plotting the convergence error as a function of the number of potential energy evaluations.}
\label{fig:NH3_convergence_psi}%
\end{figure}

Panel (a) of Fig.~\ref{fig:NH3_convergence_psi} compares the convergence rates of the integrators. 
The convergence error, defined as the distance $\lVert \psi^{(\Delta t)}_{t}-\psi^{(\Delta t/2)}_{t} \rVert$, where $\psi^{(\Delta t)}_{t}$ denotes the Gaussian wavepacket at time $t$ obtained after propagation with time step $\Delta t$, is evaluated from the relation
\begin{align}
    \lVert \psi^{(\Delta t)}_{t}-\psi^{(\Delta t/2)}_{t} \rVert&=(\lVert\psi^{(\Delta t)}_{t}\rVert^2+\lVert\psi^{(\Delta t/2)}_{t}\rVert^2\nonumber\\
    &~~~-2\,\text{Re}\langle\psi^{(\Delta t)}_{t},\psi^{(\Delta t/2)}_{t}\rangle)^{1/2}.
    \label{eq:conv_err}
\end{align}
All integrators respect the theoretically predicted asymptotic order of convergence, indicated by the grey straight lines. The plateaus indicate the machine precision error, which is, for the norm, of the order of $10^{-7}$ even though the calculations were performed in double precision ($\sim10^{-15}$). This can be easily understood from Eq.~(\ref{eq:conv_err}), in which $\lVert \psi^{(\Delta t)}_{t}-\psi^{(\Delta t/2)}_{t} \rVert^{2}$ is evaluated with precision $\sim10^{-15}$, but taking the square root results in error $\sim10^{-7}$. 
Since the high-order methods require performing a considerable number of composition substeps during each time step $\Delta t$, higher-order convergence does not necessarily guarantee higher efficiency. Therefore, panel (b) measures efficiency by plotting the convergence error as a function of the number of potential energy evaluations. It shows that high-order geometric integrators are more efficient than both the second-order geometric integrator and the fourth-order Runge-Kutta method. 
For example, for a large error of $10^{-1}$, the fourth-order geometric integrator is approximately twice more efficient than the other integrators, while for an error of $10^{-4}$, the most efficient integrator is the eighth-order one, being ten times faster than the second-order integrator.
Thus, high-order geometric integrators can enhance accuracy and efficiency simultaneously.

As shown in Fig. S2 of the supplementary material, the increased accuracy of the wavefunction translates into improved accuracy of physical observables. Unfortunately, the advantages of high-order integrators are limited in spectroscopy applications, because the maximum allowed time step is also constrained by the spectral range, and therefore by the period of the fastest vibration in the system.

\begin{figure} 
\includegraphics{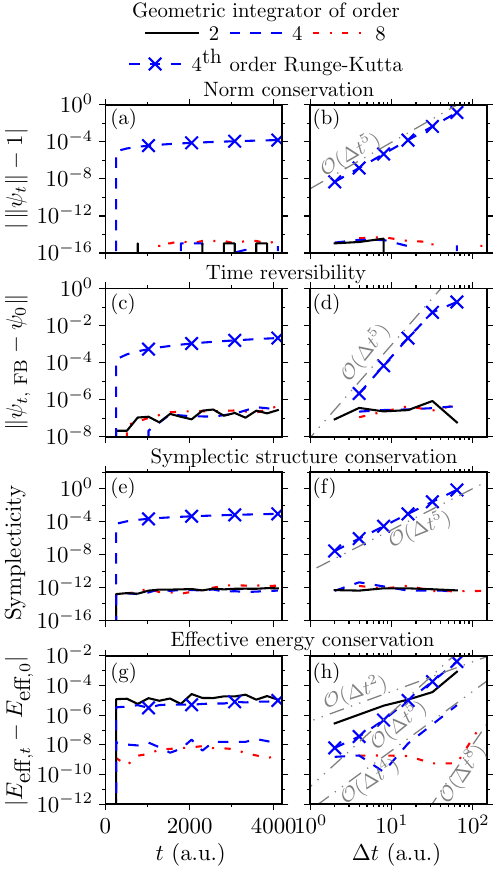}
\caption{Geometric properties of various integrators for the adiabatic single-Hessian GWD in the first excited state of ammonia as a function of time $t$ for a relatively large time step $\Delta t = 16 \, \text{a.u.}$ (left panels), and as a function of the time step $\Delta t$ measured at the final time $t_{f}=4096 \, \text{a.u.}$ (right panels). 
Norm conservation [panels (a) and (b)], time reversibility [panels (c) and (d)], and the conservation of symplectic structure [panels (e) and (f)] and effective energy [panels (g) and (h)] are shown.}
\label{fig:NH3_geom_prop}%
\end{figure}

In addition to the improved efficiency, geometric integrators introduced in Sec.~\ref{subsec:geom_integrators} preserve exactly all geometric properties of on-the-fly ab initio single-Hessian GWD, except for the conservation of the effective energy (which is conserved only approximately due to the splitting nature of the integrators).
Figure~\ref{fig:NH3_geom_prop} shows how the time reversibility and the conservation of norm, symplectic structure, and effective energy depend on time and on the time step. 
Panels (a) and (b) confirm the exact conservation of the norm by the geometric integrators. In contrast, the fourth-order Runge-Kutta method fails to conserve the norm unless it is fully converged, which requires using a very small time step. 
In panels (c) and (d), the time reversibility is checked by plotting the distance $\lVert\psi_{t,\text{FB}}-\psi_{0}\rVert$ between the initial state $\psi_{0}$ and the \textquotedblleft forward-backward\textquotedblright\ propagated state $\psi_{t,\text{FB}}$, i.e., the state propagated first forward in time for time $t$ and then backward in time for time $t$. 
Unlike the fourth-order Runge-Kutta method, geometric integrators are time-reversible. The low drift over time [panel (c)] is due to the accumulation of roundoff errors originating especially in the low precision of the electronic structure. An electronic-structure calculation of much higher precision but low accuracy would exhibit even better reversibility despite giving physically incorrect results [compare the plateaus in Figs.~\ref{fig:NH3_geom_prop}(c) and S4(c)]. 
Panels (e) and (f) of Fig.~\ref{fig:NH3_geom_prop} show that, unlike the fourth-order Runge-Kutta method, all geometric integrators conserve the non-canonical symplectic structure~(\ref{eq:sympl_QPS}) over long time, regardless of the size of the time step. For the Runge-Kutta flow, the Jacobian required in the definition~(\ref{eq:Frob_distance}) of symplecticity is constructed by combining Eq.~(\ref{eq:HJV}) with Eqs. (B32)-(B34) from Ref.~\onlinecite{Fereidani_Vanicek:2023a_v2}.
Finally, panels (g) and (h) analyze the conservation of the effective energy. Although single-Hessian GWD conserves the effective energy exactly, the geometric integrators conserve it only approximately due to their splitting nature.~\cite{book_Hairer_Wanner:2006,Roulet_Vanicek:2019} 
For nearly all time steps analyzed, the eighth-order integrator exhibits a plateau in accuracy at $10^{-8}$, a value that is constrained by the precision of the electronic structure. Figures S3 and S4 of the supplementary material show that the same integrator, employed to perform dynamics in a 20-dimensional coupled Morse potential, reaches machine precision accuracy over a wide range of time steps.

Overall, high-order geometric integrators are not only more efficient than the basic second-order algorithm but also better preserve the effective energy of the wavepacket. 
Although the Runge-Kutta method preserves none of the geometric properties, its convergence for the geometric properties appears to be $\mathcal{O}(\Delta t^{5})$ [see panels (b), (d), (f), (h)], i.e., faster than the $\mathcal{O}(\Delta t^{4})$ convergence for the wavefunction [Fig.~\ref{fig:NH3_convergence_psi}(a)].

\subsection{ $\bold{\text{CF}_{2}+e^{-}\leftarrow\text{CF}_{2}^{-}}$ photoelectron spectrum \label{subsec:CF2}}

Encouraged by the unique properties of single-Hessian GWD, we used it to compute the vibrationally resolved photoelectron spectrum of the difluorocarbene anion, a weakly anharmonic triatomic molecule.~\cite{Petrenko_Rauhut:2017,Zhang_Vanicek:2025}
We employed the optimal eighth-order integrator with a time step $\Delta t=32\, \textrm{a.u.}$, a combination that ensures full convergence of the spectrum (see Fig. S2 of the supplementary material).

Figure~\ref{fig:CF2_spectra} compares the simulated and experimental spectra.
Both global harmonic models exhibit poor peak spacing compared to local harmonic and single-Hessian approximations. The spacing of the peaks is related, via the Fourier transform~(\ref{eq:spectra}), to the recurrence time of the wavepacket autocorrelation function~(\ref{eq:auto_corr}),~\cite{Heller:1981a,book_Heller:2018,book_Tannor:2007} which is mainly determined by the trajectory guiding the Gaussian wavepacket~(\ref{eq:HGWP}). This explains the observed results because, while harmonic models fail to describe the anharmonic final-state surface of difluorocarbene, the local harmonic and single-Hessian Gaussian wavepackets follow the same exact anharmonic classical guiding trajectory.

With the exception of the adiabatic harmonic model, all methods capture the correct spectral envelope. This feature of the spectrum is determined by the initial decay of the autocorrelation function, i.e., by the short-time dynamics of the wavepacket.~\cite{Heller:1981a,book_Heller:2018,book_Tannor:2007}
In the adiabatic harmonic model, constructed about the equilibrium geometry of the PES of the final electronic state, the initial decay is too fast, resulting in a broader spectrum.
In contrast, all other approximations accurately represent the Franck-Condon region of the PES, which determines the spectral envelope.

We further note that the peak widths are very similar in the local harmonic and all three single-Hessian spectra. This aspect of the spectrum reflects the decay of recurrences in the autocorrelation function.~\cite{Heller:1981a,book_Heller:2018,book_Tannor:2007}
Because these recurrences are primarily controlled by the trajectory guiding the wavepacket, they remain very similar in all four cases. The contribution from the dynamics of the width of the wavepacket is minor and further attenuated by the damping function applied before performing the Fourier transform (see Sec.~\ref{sec:details}).

\begin{figure*} 
\includegraphics{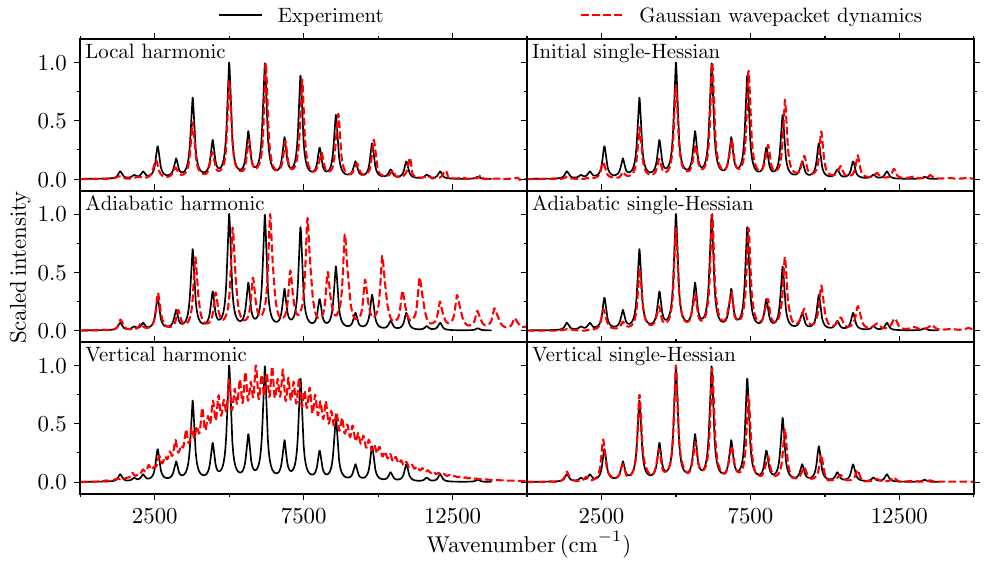}
\caption{Photoelectron spectrum of $\text{CF}_{2}^{-}$. Spectra computed from the local harmonic, adiabatic and vertical harmonic, and initial, adiabatic, and vertical single-Hessian GWD are compared to the experimental spectrum.~\cite{Murray_Lineberger:1988}}
\label{fig:CF2_spectra}%
\end{figure*}


\subsection{$\bold{S_{1} \leftarrow S_{0}}$ absorption spectrum of methylamine \label{subsec:MAM}}

To explore the limitations of single-Hessian GWD in the simulation of spectra of polyatomic molecules, we computed the vibrationally resolved absorption spectrum of methylamine.~\cite{Kletnieks_Vanicek:2026a}
Torsional motion around the CN bond makes this molecule highly anharmonic, challenging the accuracy of single-trajectory semiclassical methods.~\cite{Begusic_Vanicek:2022}
The single-Hessian spectra were obtained by post-processing the local harmonic GWD trajectory from Ref.~\onlinecite{Kletnieks_Vanicek:2026a}, where the second-order VTV integrator with a time step $\Delta t=10\, \textrm{a.u.}$ was used.

\begin{figure*} 
\includegraphics{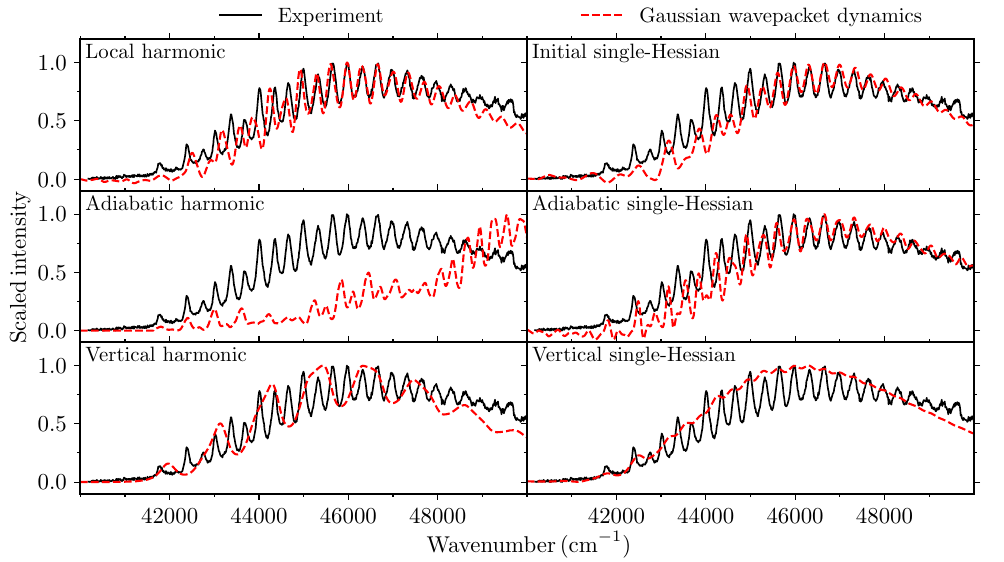}
\caption{Absorption spectrum of methylamine. Spectra computed from the local harmonic, adiabatic and vertical harmonic, and initial, adiabatic, and vertical single-Hessian GWD are compared to the experimental spectrum.~\cite{Hubin-Franskin_Jones:2002}}
\label{fig:MAM_spectra}%
\end{figure*}

Figure~\ref{fig:MAM_spectra} compares the simulated and experimental spectra.
The adiabatic harmonic spectrum is extremely broad ($\approx 100000\, \text{cm}^{-1}$) due to a rapid initial decay of the wavepacket autocorrelation function.~\cite{Kletnieks_Vanicek:2026a}
Notably, the adiabatic single-Hessian GWD--based on the same Hessian calculation--reproduces the experimental spectral envelope accurately. This result reflects the dominant role of the guiding trajectory, rather than the evolution of the width, in determining the short-time dynamics of the wavepacket.
Whereas the vertical harmonic model shows an incorrect spacing of the peaks, the underlying anharmonic trajectory enables the local harmonic and all single-Hessian GWD methods to describe the peak spacing accurately.
In contrast to the spectrum of difluorocarbene, the widths of the peaks in the vertical single-Hessian spectrum differ substantially from those in the adiabatic and initial single-Hessian spectra. This behavior stems from the presence of an imaginary frequency in the vertical Hessian, which causes the vertical single-Hessian Gaussian wavepacket to spread rapidly, leading to a fast decay of recurrences in the autocorrelation function and incorrectly broadened peaks.
Finally, the small, yet visible, nonphysical negative intensities in the local harmonic and single-Hessian spectra are due to the nonlinear character of their effective Hamiltonians~(\ref{eq:Heff}).~\cite{Vanicek:2023}

\section{Conclusion \label{sec:conclusion}}

In our comprehensive analysis, we have demonstrated that, in contrast to the popular local harmonic GWD, the single-Hessian GWD is symplectic, conserves the effective energy, and approximately conserves the exact energy in bounded regions of the potential. 
Furthermore, our numerical simulations suggest that, while computationally cheaper than the original local harmonic method, single-Hessian GWD maintains the same $\mathcal{O}(\hbar)$ error of observables with respect to the semiclassical parameter $\hbar$.

In the numerical examples, we have investigated diatomic molecules with varying degrees of anharmonicity. Spectra obtained from the single-Hessian GWD consistently outperformed those from global harmonic models--still widely used in computational chemistry--and achieved accuracy comparable to perturbation theory, the standard correction to harmonic models implemented in some commercial software packages. In general, perturbation theory and single-Hessian GWD can offer complementary information since they include anharmonicity in a static and dynamic manner, respectively. Whereas perturbation theory captures peak positions more accurately, single-Hessian GWD provides a better description of peak intensities.

However, the primary domain of applicability of single-Hessian GWD lies in weakly anharmonic polyatomic molecules. The spectra of such systems typically exhibit only partial vibrational resolution due to external broadening. In these cases, where highly precise position of the peaks is not required, on-the-fly ab initio single-Hessian GWD provides a computationally efficient and sufficiently accurate description.
Therefore, we have introduced high-order geometric integrators that enhance the accuracy and efficiency of on-the-fly ab initio simulations without compromising the geometric properties of the method. These integrators allow for machine-precision accuracy even for large time steps, facilitating the modeling of larger molecules using advanced electronic structure methods.

Then, we have applied the single-Hessian approximation to compute vibrationally resolved electronic spectra of two polyatomic molecules, difluorocarbene and methylamine.
Our numerical results have confirmed the accuracy of the single-Hessian GWD in simulating low- and medium-resolution spectra. The method has successfully reproduced the experimental spectra, outperforming global harmonic models and matching the accuracy of local harmonic GWD but at a much lower computational expense.
Yet, as shown in the simulation of methylamine and corroborated by earlier studies,~\cite{Begusic_Vanicek:2022} the choice of the reference Hessian can influence the results of single-Hessian GWD. Although the guiding trajectory generally plays a more dominant role than the evolution of the width, the presence of an imaginary frequency in the reference Hessian can suppress revivals of the autocorrelation function. In such cases, knowledge of the local features of the PES can provide partial guidance in selecting the optimal reference Hessian.

Single-Hessian GWD is subject to well-known limitations: (i)~being accurate only for short times, this method cannot describe high-resolution spectra and (ii)~being based on a single classical trajectory, this method cannot describe tunnelling splittings and spectral effects arising from wavepacket splitting and interference between components of the wavepacket.

Finally, the single-Hessian GWD is constrained to treat initial states that are purely Gaussian in shape, hence it is not suitable to simulate, e.g., single vibronic level~\cite{Tapavicza:2019,Zhang_Vanicek:2025} or Herzberg-Teller spectra,~\cite{Patoz_Vanicek:2018} which require propagating non-Gaussian wavepackets.~\cite{Vanicek_Zhang:2025_v2}
To accelerate calculations of such spectra, we are currently exploring a combination of single-Hessian approximation with Hagedorn wavepacket dynamics.~\cite{Hagedorn:1998,Lasser_Lubich:2020,Vanicek_Zhang:2025_v2}


\section*{Supplementary material}
See the supplementary material for (i)~comparison between the numerical and analytical evolution of the width of a single-Hessian Gaussian wavepacket, (ii)~convergence of observables obtained using different integrators for the single-Hessian GWD in the first excited state of ammonia, (iii)~analysis of single-Hessian GWD in a twenty-dimensional coupled Morse potential, (iv)~parameters of the diatomic molecules, and (v)~optimized geometries, harmonic frequencies, and spectral shifts of ammonia, difluorocarbene, and methylamine.

\section*{Acknowledgments}
The authors thank Fabian Kr\"{o}ninger for discussions.
This research was supported by the Swiss National Science Foundation with Grant No. 10005187 and by the EPFL. 

\section*{Data availability}
The data that support the findings of this study are openly available in Zenodo at http://doi.org/10.5281/zenodo.18609639.

\appendix


\section{Single-Hessian GWD in one-dimensional systems \label{sec:param}}

The authors of Ref.~\onlinecite{Arickx_VanLeuven:1986} showed that one-dimensional variational GWD could be mapped to the classical dynamics of a particle in a two-dimensional configuration space. To prove this statement, they introduced a new parametrization $(q_{t},\,p_{t},\,w_{t},\,u_{t})$ of the Gaussian~(\ref{eq:GWP}), where 
\begin{align}
    \mathcal{A}_{t} & = u_{t}/w_{t},\label{eq:U}\\
    \hbar/(2\mathcal{B}_{t}) & = w_{t}^{2}.\label{eq:W}
\end{align}
In this parametrization, $w_{t}$ is the spread of the wavepacket, measured by the root mean square deviation, i.e., the square root of the position variance.

Here, we show that such a statement holds for single-Hessian GWD as well. To justify it, we express the Hamiltonian function~(\ref{eq:H_map2}) as a sum of two terms, $\bar{h}_{\text{SHA}}(z_{t})= h_{\text{cl}}(p_{t},q_{t})+h_{\text{sc}}(\mathcal{A}_{t},\mathcal{B}_{t})$,~\cite{Begusic_Vanicek:2019} where $h_{\text{cl}}= T(p_{t})+V(q_{t})$ is the classical Hamiltonian, for which $(q_{t},p_{t})$ are canonically conjugate variables, and
\begin{align}
    h_{\text{sc}} & = (\hbar/4)[\mathcal{A}_{t}^{2}/(m\mathcal{B}_{t}) + \mathcal{B}_{t}/m+ \kappa_{\text{ref}}/\mathcal{B}_{t}] \nonumber\\
    &=u_{t}^{2}/(2m) + \hbar^{2}/(8mw_{t}^{2})+\kappa_{\text{ref}} w_{t}^{2}/2
    \label{eq:H_map3}
\end{align}
is the semiclassical contribution, and where we substituted from Eqs.~(\ref{eq:U}) and~(\ref{eq:W}) in the second line.
Parameters $u_{t}$ and $w_{t}$ are canonically conjugate variables for the Hamiltonian~(\ref{eq:H_map3}) and evolve according to canonical Hamilton's equations of motion
\begin{align}
    \dot{w}_{t} & =\frac{\partial h_{\text{sc}}}{\partial u_{t}}=u_{t}/m,\label{eq:WEOM_SHA}\\
    \dot{u}_{t} & =-\frac{\partial h_{\text{sc}}}{\partial w_{t}}=\hbar^{2}/(4mw_{t}^{3})-\kappa_{\text{ref}} w_{t}.\label{eq:UEOM_SHA}
\end{align}
Thus, single-Hessian GWD in one spatial dimension is exactly represented by classical dynamics of a particle in a two dimensional configuration space (or, equivalently, a four-dimensional phase space).


\section{Explicit form of $i_{X_{h_{\text{SHA}}}}$ and derivation of the 1-form $dh_{\text{SHA}}$\label{sec:1-form}}

For a Gaussian wavepacket with parameters $q,p,\mathcal{A},\mathcal{B},\phi,\text{and}\,\delta$, symplectic form~(\ref{eq:symp_form}), and vector field 
\begin{equation}
    X_{h_{\text{SHA}}}=\dot{q}_{j}\frac{\partial}{\partial q_{j}}+\dot{p}_{j}\frac{\partial}{\partial p_{j}}+\dot{\mathcal{A}}_{jk}\frac{\partial}{\partial \mathcal{A}_{jk}}+\dot{\mathcal{B}}_{jk}\frac{\partial}{\partial \mathcal{B}_{jk}}+\dot{\phi}\frac{\partial}{\partial\phi}+\dot{\delta}\frac{\partial}{\partial\delta},
\end{equation}
the interior product $i_{X_{h_{\text{SHA}}}}\omega(\cdot)=\omega(X_{h_{\text{SHA}}},\cdot)$ gives~\cite{Ohsawa_Leok:2013}
\begin{align}
    i_{X_{h_{\text{SHA}}}}\omega&=\mathcal{N}(\mathcal{B},\delta)\left\{\dot{q}^{T}\cdot dp - \dot{p}^{T}\cdot dq\right.\nonumber\\
    &~~~- p^{T}\cdot[\dot{q}\,\text{Tr}(\mathcal{B}^{-1}\cdot d\mathcal{B})-dq\,\text{Tr}(\mathcal{B}^{-1}\dot{\mathcal{B}})]/2  \nonumber\\
     &~~~ - (2/\hbar)p^{T}\cdot(\dot{q}\,d\delta-dq\,\dot{\delta})\nonumber\\
     &~~~+(\hbar/8)[\text{Tr}(\mathcal{B}^{-1}\cdot\dot{\mathcal{A}})\text{Tr}(\mathcal{B}^{-1}\cdot d\mathcal{B})\nonumber\\
     &~~~-\text{Tr}(\mathcal{B}^{-1}\cdot\dot{\mathcal{B}})\text{Tr}(\mathcal{B}^{-1}\cdot d\mathcal{A})]\nonumber\\
      &~~~+(\hbar/4)\text{Tr}(\mathcal{B}^{-1}\cdot\dot{\mathcal{A}}\cdot\mathcal{B}^{-1}\cdot d\mathcal{B}\nonumber\\
      &~~~-\mathcal{B}^{-1}\cdot d\mathcal{A}\cdot\mathcal{B}^{-1}\cdot\dot{\mathcal{B}})  \nonumber\\
    &~~~+[\text{Tr}(\mathcal{B}^{-1}\cdot\dot{\mathcal{A}})d\delta-\text{Tr}(\mathcal{B}^{-1}\cdot d\mathcal{A})\dot{\delta}\nonumber\\
    &~~~-\text{Tr}(\mathcal{B}^{-1}\cdot\dot{\mathcal{B}})d\phi+\text{Tr}(\mathcal{B}^{-1}\cdot d\mathcal{B})\dot{\phi}]/2 \nonumber\\
     &~~~\left. +(2/\hbar)(\dot{\phi}\,d\delta-\dot{\delta}\,d\phi)\right\}.
\end{align}

Following the approach used by Ohsawa and Leok for the variational GWD,~\cite{Ohsawa_Leok:2013} one can compute the 1-form $dh_{\text{SHA}}$ needed in Sec.~\ref{subsec:symplectic_structure} using the relations
\begin{align}
\partial \bar{h}_{\text{SHA}}/\partial q &  =V^{\prime}(q),\label{eq:V_q}\\
\partial \bar{h}_{\text{SHA}}/\partial p &  =m^{-1}\cdot{p},\label{eq:T_p}\\
\partial \bar{h}_{\text{SHA}}/\partial\mathcal{A} &  = (\hbar/4)\,(m^{-1}\cdot\mathcal{A}\cdot\mathcal{B}%
^{-1}+\mathcal{B}^{-1}\cdot\mathcal{A}\cdot m^{-1}), \label{eq:T_A}\\
\partial \bar{h}_{\text{SHA}}/\partial\mathcal{B} &  =(\hbar
/4)\,[m^{-1}-\mathcal{B}^{-1}\nonumber\\
&~~~\cdot(\mathcal{A}\cdot m^{-1}\cdot\mathcal{A}+\kappa_{\text{ref}})\cdot\mathcal{B}^{-1}],\label{eq:TV_B}\\
\partial \bar{h}_{\text{SHA}}/\partial \phi&=\partial \bar{h}_{\text{SHA}}/\partial \delta=0
\end{align}
for the derivatives of the Hamiltonian function $\bar{h}_{\text{SHA}}$ [Eq.~(\ref{eq:H_map2})] with respect to the parameters~(\ref{eq:local_coord}) of the Gaussian wavepacket~(\ref{eq:GWP}). As a result, we have
\begin{align}
dh_{\text{SHA}}&=\mathcal{N}(\mathcal{B},\delta
)\,[V^{\prime}(q)^{T}\cdot dq+p^{T}\cdot m^{-1}\cdot d{p}\nonumber\\
&~~~+\text{Tr}(\epsilon \cdot d\mathcal{A})+ \text{Tr}(\zeta\cdot d\mathcal{B})-(2/\hbar)\,\bar{h}_{\text{SHA}}d\mathcal{\delta}],\label{eq:grad_E}%
\end{align}
where $\epsilon=\partial \bar{h}_{\text{SHA}}/\partial\mathcal{A}$ and $\zeta=\partial \bar{h}_{\text{SHA}}/\partial\mathcal{B}-\bar{h}_{\text{SHA}}\mathcal{B}^{-1}/2$.


\bibliographystyle{aipnum4-2}

\bibliography{SH-GWD_v38}

\end{document}


\title{Supplementary material for ``On the single-Hessian Gaussian wavepacket dynamics''}

\author{Davide Barbiero}
\author{Ji\v{r}\'i J. L. Van\'i\v{c}ek}
\email{jiri.vanicek@epfl.ch}

\affiliation{Laboratory of Theoretical Physical Chemistry, Institut des Sciences et Ing\'enierie Chimiques, Ecole Polytechnique F\'ed\'erale de Lausanne (EPFL), CH-1015, Lausanne, Switzerland}

\date{\today}

\begin{abstract}
This document, which provides supplementary information to the main text, contains (a)~comparison between the numerical and analytical evolutions of the width of a single-Hessian Gaussian wavepacket, (b)~ analysis of the convergence of observables obtained using different integrators for the single-Hessian GWD in the first excited state of ammonia, (c)~corresponding analysis of single-Hessian GWD in a twenty-dimensional coupled Morse potential dynamics, (d) parameters of the diatomic molecules, and (e)~details of the ab initio simulations of ammonia, difluorocarbene, and methylamine.
\end{abstract}

\maketitle

\section{Exact propagation of the width of a single-Hessian Gaussian wavepacket}

\begin{figure} [h]
\includegraphics{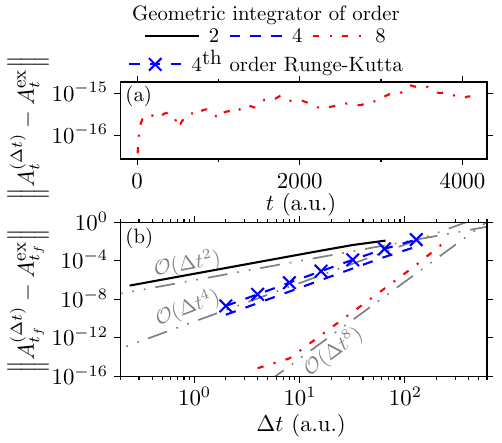}
\caption{Dynamics of the width $A_{t}\equiv P_{t}\cdot Q_{t}^{-1}$ of a single-Hessian Gaussian wavepacket in the first excited state of ammonia. (a) Distance between numerically propagated (but fully converged) width matrix $A_{t}^{(\Delta t)}$ and exact, analytically propagated width matrix $A_{t}^{\text{ex}}$ as a function of time. (b) Convergence of the geometric integrators and of the fourth-order Runge-Kutta method. Convergence is measured as the distance between the numerically propagated ($A_{t_{f}}^{(\Delta t)}$) and analytically evaluated ($A_{t_{f}}^{\text{ex}}$) width matrices at the final time $t_{f} =4096 \, \text{a.u.}$ as a function of the time step $\Delta t$.}
\label{fig:Morse_width}
\end{figure}

The decoupling between the evolutions of the center $(q_{t},p_{t})$ and of the width $(Q_{t},P_{t})$ of a single-Hessian Gaussian wavepacket allows combining numerical propagation of the center with explicit, exact propagation of the width matrix.
Panel (a) of Fig.~\ref{fig:Morse_width} shows that the numerically propagated (using eighth-order integrator and $\Delta t=4 \, \text{a.u.}$, which guarantees full convergence) and analytically propagated width matrices agree to machine precision as a function of time. This validates the implementation of the exact propagation scheme from Sec.~IIIA of the main text.
Panel (b) shows that the numerical integrators require taking many steps of decreasing size $\Delta t$ for convergence, whereas the exact propagation gives the exact calculation in one step 
and does not have to lower the error by decreasing $\Delta t$.

\section{Convergence of observables}

\begin{figure} [h]
\includegraphics{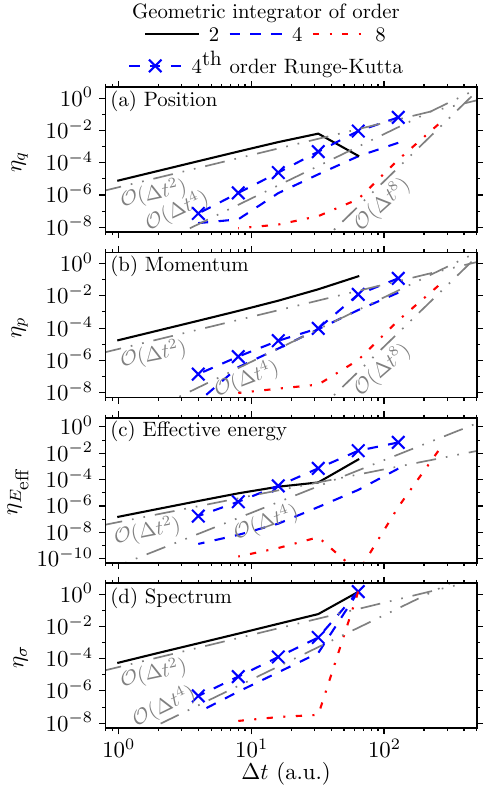}
\caption{Convergence of observables obtained from geometric integrators and from the fourth-order Runge-Kutta method for the adiabatic single-Hessian GWD in the first excited state of ammonia. For the expectation values of (a) position $q$, (b) momentum $p$, and (c) effective energy $E_{\text{eff}}$, convergence is measured by the error $\eta_{A}=|(A_{t_{f}}^{(\Delta t)}-A_{t_{f}}^{(\Delta t/2)})/A_{t_{f}}^{(\Delta t/2)}|$ as a function of the time step $\Delta t$, where $A_{t_{f}}^{(\Delta t)}$ denotes the property at the final time $t_{f}=4096 \, \textrm{a.u.}$ obtained after propagation with time step $\Delta t$. In panel (d), the convergence error of the spectrum $\sigma$ is measured using Eq.~(\ref{eq:dist_spec}).}
\label{fig:convergence_obs}%
\end{figure}

In Fig. 5 of the main text, we showed that high-order geometric integrators can simultaneously enhance accuracy and efficiency of simulations by measuring the convergence error of the wavefunction. Here, we show that the increased accuracy of the wavefunction translates into improved accuracy of physical observables.
Panels (a) and (b) of Fig.~\ref{fig:convergence_obs} compare the convergence rates of the integrators for the expectation values of position and momentum, which in the single-Hessian GWD coincide with their classical values $q_{t}$, $p_{t}$. Again, all integrators follow the predicted asymptotic order of convergence.
Convergence of the effective energy is plotted in panel (c). While the effective energy is not a physical observable, its convergence behavior is expected to resemble that of the exact energy, which cannot be evaluated analytically and which we therefore did not compute due to the high cost and potential inaccuracies associated with its numerical evaluation. 
Finally, panel (d) shows the convergence of the vibrationally resolved electronic spectrum of ammonia. Although the unbroadened spectrum is generally a mathematical distribution, when convolved with a smooth broadening function, it becomes a smooth function,~\cite{Hormander:1990} and its convergence error can be evaluated as the $L^{2}$-distance
\begin{equation}
    \eta_{\sigma}=\lVert\sigma^{(\Delta t)}-\sigma^{(\Delta t/2)}\rVert/\lVert\sigma^{(\Delta t/2)}\rVert,\label{eq:dist_spec}
\end{equation}
where $\sigma^{(\Delta t)}$ is the spectrum obtained from the autocorrelation function [Eq. (2) of the main text] evaluated using a time step $\Delta t$ and $\sigma^{(\Delta t/2)}$ is the spectrum obtained from the autocorrelation function evaluated every two steps of size $\Delta t/2$.
Computing $\sigma^{(\Delta t/2)}$ from the autocorrelation function evaluated only once after every two steps of $\Delta t/2$ ensures that any difference between the two spectra reflects only the accuracy of the propagated wavefunction and not the number of sampling points in the autocorrelation function.
For a large time step $\Delta t = 64 \, \textrm{a.u.}$, the spectrum is inaccurate regardless of the integrator because this time step is insufficient to resolve the entire frequency range. As the time step decreases, the accuracy of the eighth-order integrator immediately reaches the plateau due to machine precision, whereas the lower-order integrators follow the expected asymptotic behavior.

\section{Dynamics in a twenty-dimensional coupled Morse potential}

\begin{figure} [h]
\includegraphics{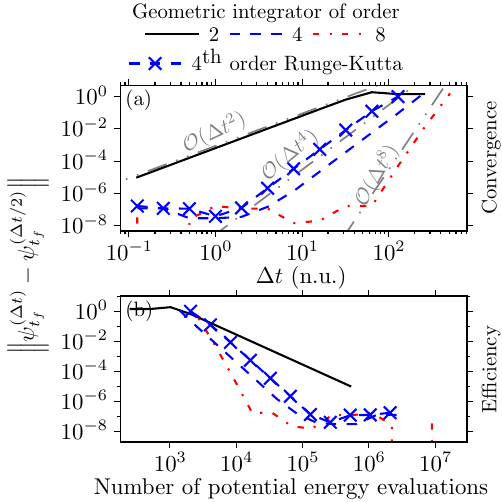}
\caption{Convergence and efficiency of the geometric integrators and of the fourth-order Runge-Kutta method for the adiabatic single-Hessian GWD in a twenty-dimensional coupled Morse potential~(\ref{eq:Coupled-Morse}). Convergence [panel (a)] is measured by the convergence error~[Eq.~(55) of the main text] at the final time $t_{f} =65536 \, \text{n.u.}$ as a function of the time step $\Delta t$. 
Efficiency [panel (b)] is measured by plotting the convergence error as a function of the number of potential energy evaluations. }
\label{fig:Morse_convergence}
\end{figure}

\begin{figure} [h]
\includegraphics{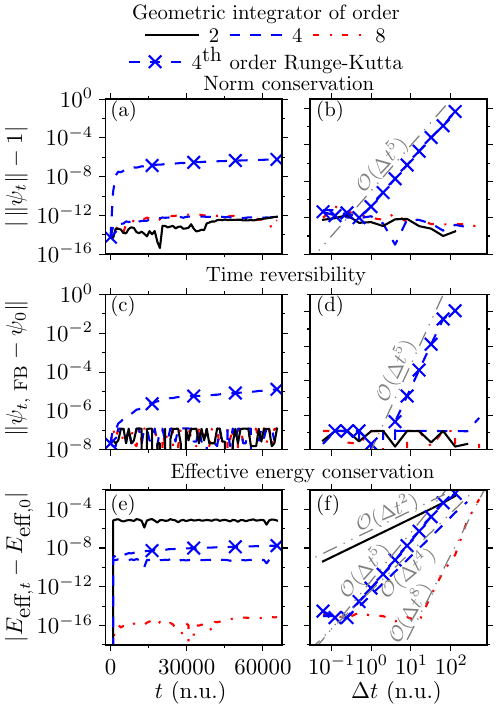}
\caption{Geometric properties of various integrators for the adiabatic single-Hessian GWD in a twenty-dimensional coupled Morse potential as a function of time $t$ for a relatively large time step $\Delta t = 8 \, \text{n.u.}$ (left panels), and as a function of the time step $\Delta t$ measured at the final time $t_{f} = 65536 \, \text{n.u.}$ (right panels). 
Norm conservation [panels (a) and (b)], time reversibility [panels (c) and (d)], and effective energy conservation [panels (e) and (f)] are shown.}
\label{fig:Morse_geom_prop}
\end{figure}

Figures~\ref{fig:Morse_convergence} and~\ref{fig:Morse_geom_prop} demonstrate the advantages of the geometric integrators over the fourth-order Runge-Kutta method when applied to the single-Hessian GWD in a twenty-dimensional coupled Morse potential 
\begin{equation}
V(q)=V_{\text{cpl}}(q)+\sum_{j=1}^{20}V_{j}(q_{j}).\label{eq:Coupled-Morse}%
\end{equation}
Potential~(\ref{eq:Coupled-Morse}) is composed of twenty one-dimensional Morse potentials 
\begin{equation}
V_{j}(q_{j}):=D_{e}^{\prime}\,\big\{1-\exp\big[-a_{j}^{\prime}\,(q_{j}-q_{\text{eq}%
,j})\big]\big\}^{2},%
\label{eq:1D_Morse}%
\end{equation}
with the same equilibrium position, $q_{\text{eq},j}=10$, and dissociation energy, $D_{e}^{\prime}=0.1$, but with different decay parameters, $a_{j}^{\prime}=\chi_{j}^{\prime}(8D_{e}^{\prime})^{1/2}$, and dimensionless anharmonicity parameters $\chi_{j}^{\prime}\,\,(j=1,\dots,20)$ varying uniformly in the range between $0.001$ and $0.005$.
The coupling term
\begin{equation}
V_{\text{cpl}}(q):=D_{e}\,\big\{1-\exp\big[-a^{T}\cdot(q-q_{\text{eq}})\big]\big\}^{2}\label{eq:cpl_Morse}%
\end{equation}
was defined with parameters $D_{e}=0.075$, $a=(8D_{e})^{1/2}\chi$, $\chi^{T}=(\chi_{1},\dots,\chi_{20})$, and $\chi_{j}=(3/4)^{1/2}\,\chi_{j}^{\prime}$. The twenty-dimensional initial Gaussian wavepacket had zero position, zero momentum, and a diagonal width matrix with non-zero elements $Q_{jj}=(4\,d_{e}^{\prime}\chi_{j}^{\prime})^{-1/2}$, $P_{jj}=(4\,d_{e}^{\prime}\chi_{j}^{\prime})^{1/2}\,i$.
All values are presented in natural units (n.u.) with $m = 1$ and $\hbar=1$.

The convergence rates and efficiency of different integrators are compared in Fig.~\ref{fig:Morse_convergence}. Panel (a) confirms the theoretical asymptotic order of convergence of the integrators, and the ability of high-order methods to obtain converged results with much larger time steps than the standard second-order integrator. 
Panel (b) shows the superior efficiency of high-order geometric integrators. For example, for convergence errors below a rather large error of $10^{-2}$, the eighth-order integrator is already more efficient than the fourth- and second-order integrators. 

Beside their improved efficiency, the advantage of geometric integrators is their ability to preserve the geometric properties of the single-Hessian GWD, as shown in Fig.~\ref{fig:Morse_geom_prop}.
Panels (a) and (b) confirm the exact conservation of the norm, while panels (c) and (d) confirm the reversibility. The fourth-order Runge-Kutta calculation conserves these properties only when fully converged, which requires using a very small time step.  
Panels (e) and (f) show that, due to their splitting nature, even the geometric integrators conserve the effective energy only approximately and better conservation requires a smaller time step. The precise analytical expression of the potential explains the much lower plateau for the convergence errors compared to Fig.~6 of the main text. 
In panel (e), the fourth-order Runge-Kutta integrator shows a gradual drift in the energy over time. In contrast, no such drift is observed for the geometric integrators. Ideally, the total propagation time should be increased to further examine this advantageous property of geometric integrators.

We do not repeat the analysis of symplecticity, which was conducted for ammonia in Fig.~6 of the main text, because its evaluation for the twenty-dimensional system is unnecessarily expensive. 


\section{Absorption spectra of diatomic molecules}

Table~\ref{tab:geom_diat} reports the parameters of the four diatomic molecules studied in Sec. VA of the main text. 

\begin{table}[H]
    \caption{Experimental parameters of CuI, BaO, BO, and BiCl.~\cite{Huber_Herzberg:1979} The vibrational wavenumbers in the ground ($\tilde{\nu}^{\prime\prime}$) and excited ($\tilde{\nu}^{\prime}$) electronic states, the anharmonicity parameter $\chi$ of the excited-state surface, and the Huang-Rhys parameter $S$ between the two surfaces are reported.}
    \label{tab:geom_diat}
    \centering
    \begin{tabular}{ccdddd}
        \hline \hline
Molecule & Transition & \multicolumn{1}{c}{$\tilde{\nu}^{\prime\prime}/\text{cm}^{-1}$} & \multicolumn{1}{c}{$\tilde{\nu}^{\prime}/\text{cm}^{-1}$} & \multicolumn{1}{c}{$\chi$} & \multicolumn{1}{c}{$S$} \\
\hline
CuI & $\tilde{\text{C}}\,^1\Sigma^{+}\leftarrow\tilde{\text{X}}\,^1\Sigma^{+}$ & 264.5 & 229.7 & 0.0023 & 1.2787 \\
BaO & $\tilde{\text{A}}\,^1\Sigma^{+}\leftarrow\tilde{\text{X}}\,^1\Sigma^{+}$ & 669.8 & 499.7 & 0.0033 & 5.3674 \\
BO & $\tilde{\text{B}}\,^2\Sigma^{+}\leftarrow\tilde{\text{X}}\,^2\Sigma^{+}$ & 1885.7 & 1281.7 & 0.0083 & 1.8578 \\
BiCl & $\tilde{\text{A}}\,\text{O}^{+}\leftarrow\tilde{\text{X}}\,\text{O}^{+}$ & 308.4 & 220.3 & 0.0112 & 5.0540 \\
\hline \hline
\end{tabular}
\end{table}

\section{Optimized geometries of ammonia}

Equilibrium nuclear geometries in the ground ($\text{S}_{0}$, point group $C_{3v}$) and first excited ($\text{S}_{1}$, point group $D_{3h}$) electronic states of ammonia were calculated using the CAM-B3LYP functional and the 6-31+G** basis set. The computed values are provided in Table~\ref{tab:geom_NH3} ($r$ is the NH bond length, $\theta_{1}$ is the ``umbrella'' angle, and $\theta_{2}$ is the HNH angle).

\begin{table}[H]
\caption{Equilibrium nuclear geometries in the ground and first excited electronic states of ammonia at the (TD-)CAM-B3LYP/6-31+G** level of theory.}
    \label{tab:geom_NH3}
    \centering
    \begin{tabular}{cddd}
        \hline \hline
         & \multicolumn{1}{c}{$r/\text{\r{A}}$} & \multicolumn{1}{c}{$\theta_{1}/\text{deg}$} & \multicolumn{1}{c}{$\theta_{2}/\text{deg}$}\\
         \hline 
         $\text{S}_{0}(\tilde{\text{X}}\,^{1}\text{A}_{2}^{\prime\prime})$ & 1.0136 & 20.22 & 108.71 \\
         $\text{S}_{1}(\tilde{\text{A}}\,^{1}\text{A}_{1}^{\prime})$ & 1.0492 & 0.00 & 120.00 \\
         \hline \hline
    \end{tabular}
\end{table}


\section{$\bold{\text{CF}_{2}+e^{-}\leftarrow\text{CF}_{2}^{-}}$ photoelectron spectrum}

Equilibrium nuclear geometries in the ground electronic states of $\text{CF}_{2}^{-}$ (point group $C_{2v}$) and $\text{CF}_{2}$ (point group $C_{2v}$) were calculated using the PBE0 functional and the aug-cc-pVTZ basis set. The computed values are provided in Table~\ref{tab:geom_CF2} ($r$ is the CF bond length, $\theta$ is the FCF angle). Table~\ref{tab:freq_CF2} presents the harmonic vibrational wavenumbers and the dimensionless Huang-Rhys displacement parameter $S:=\pi mc \tilde{\nu}^{\prime\prime}(\Delta q)^{2}/\hbar$ along the normal-mode coordinates. The shifts applied to the computed spectra are reported in Table~\ref{tab:shift_CF2}.

\begin{table}[H]
    \caption{Equilibrium geometries of $\text{CF}_{2}^{-}$ and $\text{CF}_{2}$ at the PBE0/aug-cc-pVTZ level of theory.}
    \label{tab:geom_CF2}
    \centering
    \begin{tabular}{cdd}
        \hline \hline
         &  \multicolumn{1}{c}{$r/\text{\r{A}}$} & \multicolumn{1}{c}{$\theta/\text{deg}$} \\
         \hline 
         $\text{CF}_{2}^{-}(\tilde{\text{X}}\,^{2}\text{B}_{1})$ & 1.4259 & 100.46 \\
         $\text{CF}_{2}(\tilde{\text{X}}\,^{1}\text{A}_{1})$ & 1.2966 & 104.75 \\
         \hline \hline
    \end{tabular}
\end{table}

\begin{table}[H]
  \caption{Computed vibrational wavenumbers of $\text{CF}_{2}^{-}$ ($\tilde{\nu}_{j}^{\prime\prime}$) and $\text{CF}_{2}$ ($\tilde{\nu}_{j}^{\prime}$), and Huang-Rhys parameters $S_{j}$ between these two species at the PBE0/aug-cc-pVTZ level of theory.}
    \label{tab:freq_CF2}
    \centering
    \begin{tabular}{ccddd}
        \hline \hline
        Mode label $j$ & Symmetry & \multicolumn{1}{c}{$\tilde{\nu}_{j}^{\prime\prime}/\text{cm}^{-1}$} & \multicolumn{1}{c}{$\tilde{\nu}_{j}^{\prime}/\text{cm}^{-1}$} & \multicolumn{1}{c}{$S_{j}$}\\
        \hline 
        1 & a${_1}$ & 901.90 & 1255.11 & 3.01 \\
        2 & a${_1}$ & 503.24 & 680.65 & 0.66 \\
        3 & b${_2}$ & 750.41 & 1150.60 & 0 \\
        \hline \hline
    \end{tabular}
\end{table}

\begin{table}[H]
    \caption{Energy shifts applied to the computed photoelectron spectra of $\text{CF}_{2}^{-}$.}
    \label{tab:shift_CF2}
    \centering
    \begin{tabular}{ld}
        \hline \hline
         Variant of GWD & \multicolumn{1}{c}{Energy shift / $\text{cm}^{-1}$} \\
         \hline 
         Adiabatic harmonic & -2528 \\
         Vertical harmonic & -1270\\
         Adiabatic single-Hessian & -1356\\
         Vertical single-Hessian & -2062 \\
         Initial single-Hessian & -891 \\
         Local harmonic & -1287 \\
         \hline \hline
    \end{tabular}
\end{table}


\section{Absorption spectrum of methylamine}

Equilibrium nuclear geometries in the ground ($\text{S}_{0}$, point group $C_{s}$) and first excited ($\text{S}_{1}$, point group $C_{1}$) electronic states of methylamine were calculated using the MS-CASPT2(4e,6o)/6-31++G** level of theory. The optimized structures in Cartesian coordinates are provided in Tables~\ref{tab:ground_MAM} and~\ref{tab:excited_MAM}. Table~\ref{tab:freq_MAM} presents the harmonic vibrational wavenumbers, while the energy shifts applied to the computed spectra are reported in Table~\ref{tab:shift_MAM}.

\begin{table}[H]
    \caption{Ground-state equilibrium geometry (in $\text{\r{A}}$) of methylamine at the MS-CASPT2(4e,6o)/6-31++G** level of theory.}
    \label{tab:ground_MAM}
    \centering
    \begin{tabular}{cddd}
         \hline \hline
         &  \multicolumn{1}{c}{$x$} & \multicolumn{1}{c}{$y$} & \multicolumn{1}{c}{$z$}\\
         \hline 
 C &          0.0137 &        0.0000 &       -0.7403\\
 H &         -0.5038 &       0.8779   &    -1.1228\\
 H &          1.0307 &       0.0000   &    -1.1460\\
 H &         -0.5038 &      -0.8779   &    -1.1228\\
 N &         -0.0701 &       0.0000   &     0.7196\\
 H &          0.3938 &       0.8170   &     1.1068\\
 H &          0.3938 &      -0.8170   &     1.1068 \\
         \hline \hline
    \end{tabular}
\end{table}
\begin{table}[H]
    \caption{Excited-state equilibrium geometry (in $\text{\r{A}}$) of methylamine at the MS-CASPT2(4e,6o)/6-31++G** level of theory.}
    \label{tab:excited_MAM}
    \centering
    \begin{tabular}{cddd}
        \hline \hline
        &  \multicolumn{1}{c}{$x$} & \multicolumn{1}{c}{$y$} & \multicolumn{1}{c}{$z$}\\
         \hline 
C & -0.0110 & -0.0022 & -0.7426 \\
H & -0.5011 &  0.9023 & -1.0937 \\
H &  1.0415 &  0.1411 & -1.0596 \\
H & -0.4816 & -0.9152 & -1.0934 \\
N & -0.0016 &  0.0010 &  0.6933 \\
H &  0.0774 &  0.8926 &  1.2242 \\
H &  0.0166 & -0.8822 &  1.2374 \\
         \hline \hline
    \end{tabular}
\end{table}

\begin{table}[H]
    \caption{Computed vibrational wavenumbers in $\text{S}_{0}$ ($\tilde{\nu}_{j}^{\prime\prime}$) and $\text{S}_{1}$ ($\tilde{\nu}_{j}^{\prime}$) electronic states of methylamine, and Huang-Rhys parameters $S_{j}$ between the corresponding two surfaces at the MS-CASPT2(4e,6o)/6-31++G** level of theory.}
    \label{tab:freq_MAM}
    \centering
    \begin{tabular}{ccddd}
        \hline \hline
        Mode label $j$ & Symmetry & \multicolumn{1}{c}{$\tilde{\nu}_{j}^{\prime\prime}/\text{cm}^{-1}$} & \multicolumn{1}{c}{$\tilde{\nu}_{j}^{\prime}/\text{cm}^{-1}$} & \multicolumn{1}{c}{$S_{j}$}\\
        \hline 
        1 & a & 325.64 & 87.34 & 3.54 \\
        2 & a & 878.94 & 616.79 & 4.40 \\
        3 & a & 989.11 & 882.32 & 0.87 \\
        4 & a & 1102.09 & 1017.49 & 0.01 \\
        5 & a & 1209.03 & 1089.02 & 0.80 \\
        6 & a & 1375.02 & 1267.51 & 0.06\\
        7 & a & 1505.07 & 1411.98 & 0.10 \\
        8 & a & 1546.47 & 1478.01 & 0.05 \\
        9 & a & 1566.89 & 1494.59 & 0.02 \\
       10 & a & 1696.78 & 1495.89 & 0.00 \\
       11 & a & 3092.44 & 3056.07 & 0.25 \\
       12 & a & 3189.20 & 3104.68 & 0.03 \\ 
       13 & a & 3233.60 & 3133.61 & 0.00 \\
       14 & a & 3551.57 & 3209.36 & 6.96 \\
       15 & a & 3634.13 & 3289.04 & 8.12 \\
        \hline \hline
    \end{tabular}
\end{table}

\begin{table}[H]
    \caption{Energy shifts applied to the computed absorption spectra of methylamine.}
    \label{tab:shift_MAM}
    \centering
    \begin{tabular}{ld}
        \hline \hline
            Variant of GWD & \multicolumn{1}{c}{Energy shift / $\text{cm}^{-1}$} \\
         \hline 
         Adiabatic harmonic & 1630 \\ 
         Vertical harmonic & 1305\\
         Adiabatic single-Hessian & 1765\\
         Vertical single-Hessian & 1305 \\
         Initial single-Hessian & 1145 \\
         Local harmonic & 1715 \\
         \hline \hline
    \end{tabular}
\end{table}


\bibliographystyle{aipnum4-2}

\bibliography{SH-GWD_v38}